\documentclass[twocolumn]{aastex631}

\usepackage{amsmath}

\begin{document}

\title{The \textit{JWST} EXCELS survey: Insights into the nature of quenching at cosmic noon}

\author[0009-0004-0844-0657]{Maya Skarbinski}
\affiliation{William H. Miller III Department of Physics and Astronomy, Johns Hopkins University,
Baltimore, MD 21218, USA}

\author[0000-0001-7883-8434]{Kate Rowlands}
\affiliation{Space Telescope Science Institute, 3700 San Martin Drive, Baltimore, MD 21218, USA}
\affiliation{William H. Miller III Department of Physics and Astronomy, Johns Hopkins University, Baltimore, MD 21218, USA}

\author[0000-0002-4261-2326]{Katherine Alatalo}
\affiliation{Space Telescope Science Institute, 3700 San Martin Drive, Baltimore, MD 21218, USA}
\affiliation{William H. Miller III Department of Physics and Astronomy, Johns Hopkins University, Baltimore, MD 21218, USA}

\author[0000-0002-8956-7024]{Vivienne Wild}
\affiliation{School of Physics and Astronomy, University of St Andrews, North Haugh, St Andrews, KY16 9SS, UK}

\author[0000-0002-1482-5818]{Adam C. Carnall}
\affiliation{Institute for Astronomy, University of Edinburgh, Royal Observatory, Edinburgh, EH9 3HJ, UK}

\author[0000-0001-9328-3991]{Omar Almaini}
\affiliation{School of Physics and Astronomy, University of Nottingham, University Park, Nottingham NG7 2RD, UK}

\author{David Maltby}
\affiliation{School of Physics and Astronomy, University of Nottingham, University Park, Nottingham NG7 2RD, UK}

\author{Thomas de Lisle}
\affiliation{School of Physics and Astronomy, University of Nottingham, University Park, Nottingham NG7 2RD, UK}

\author[0000-0001-6670-6370]{Timothy Heckman}
\affiliation{William H. Miller III Department of Physics and Astronomy, Johns Hopkins University, Baltimore, MD 21218, USA}
\affiliation{School of Earth and Space Exploration, Arizona State University, Temple, AZ 85287, USA}

\author{Ryan Begley}
\affiliation{Institute for Astronomy, University of Edinburgh, Royal Observatory, Edinburgh, EH9 3HJ, UK}

\author{Fergus Cullen}
\affiliation{Institute for Astronomy, University of Edinburgh, Royal Observatory, Edinburgh, EH9 3HJ, UK}

\author[0000-0002-1404-5950]{James S. Dunlop}
\affiliation{Institute for Astronomy, University of Edinburgh, Royal Observatory, Edinburgh, EH9 3HJ, UK}

\author[0009-0006-7827-007X]{Guillaume Hewitt}
\affiliation{School of Physics and Astronomy, University of Nottingham, University Park, Nottingham NG7 2RD, UK}

\author[0000-0003-0486-5178]{Ho-Hin Leung}
\affiliation{Institute for Astronomy, University of Edinburgh, Royal Observatory, Edinburgh, EH9 3HJ, UK}

\author{Derek McLeod}
\affiliation{Institute for Astronomy, University of Edinburgh, Royal Observatory, Edinburgh, EH9 3HJ, UK}

\author{Ross McLure}
\affiliation{Institute for Astronomy, University of Edinburgh, Royal Observatory, Edinburgh, EH9 3HJ, UK}

\author[0000-0003-3191-9039]{Justin Atsushi Otter}
\affiliation{William H. Miller III Department of Physics and Astronomy, Johns Hopkins University, Baltimore, MD 21218, USA}

\author{Pallavi Patil}
\affiliation{William H. Miller III Department of Physics and Astronomy, Johns Hopkins University, Baltimore, MD 21218, USA}

\author{Andreea Petric}
\affiliation{Space Telescope Science Institute, 3700 San Martin Drive, Baltimore, MD 21218, USA}

\author{Alice E. Shapley}
\affiliation{Department of Physics \& Astronomy, University of California, 430 Portola Plaza, Los Angeles CA 90095, USA}

\author{Struan Stevenson}
\affiliation{Institute for Astronomy, University of Edinburgh, Royal Observatory, Edinburgh, EH9 3HJ, UK}

\author[0000-0001-8728-2984]{Elizabeth Taylor}
\affiliation{Institute for Astronomy, University of Edinburgh, Royal Observatory, Edinburgh, EH9 3HJ, UK}



\begin{abstract}

We study 24 massive quiescent galaxies with $\log \textrm{M}_*/\textrm{M}_\odot > 10$ at $1 < z < 3$ with \textit{JWST}/NIRSpec medium-resolution observations from the Early eXtragalactic Continuum and Emission Line Survey (EXCELS). We reconstruct their star formation histories and find that they have large bursts ($100\textrm{ M}_{\odot} \textrm{yr}^{-1} -1000 \textrm{ M}_{\odot} \textrm{yr}^{-1}$), followed by a rapid truncation of star formation. The number densities of the quenched galaxies in our sample that we predict underwent a submillimeter phase are consistent with submillimeter galaxies being the progenitors of our quenched population. The median post-starburst visibility time is $\sim600$ Myr, with more massive galaxies ($\log \textrm{M}_*/\textrm{M}_\odot > 10.7$) exhibiting shorter visibility times than lower mass galaxies. The range of quenching times -- defined as the time from the peak starburst to the time of quiescence -- found in this sample ($0.06-1.75$ Gyr) suggests multiple quenching pathways, consistent with previous studies. We do not see evidence for quenching mechanisms varying with redshift between $1<z<3$. We detect evidence for weak AGN activity in 4 out of the 8 galaxies with robust emission line detections, based on line ratio diagnostics. Our findings suggest that there are a diverse range of quenching mechanisms at cosmic noon, and support a scenario in which the primary quenching mechanisms are rapid ($<500$ Myr) following a starburst.


\end{abstract}



\section{Introduction} 
\label{sec:intro}
Constraining the physical mechanisms responsible for the quenching of star formation remains a challenge. Star formation quenching drives the bimodal color distribution in galaxies across cosmic time \citep[e.g.][]{Bell2004, Arnouts2007}. We begin to see a significant build-up of massive, quenched galaxies during the period referred to as `cosmic noon' ($1<z<3$), when many local passive galaxies are thought to have stopped forming stars \citep[e.g.][]{Fontana2004, Ilbert2013, Muzzin2013, McLeod2021}. However, the bulk of the quenched stellar mass was assembled between $z\sim1$ and $z\sim2$, with only a small fraction ($\sim$10\%) in place by $z > 2$ \citep{McLeod2021, Weaver2023}. This raises the question of whether the nature of quenching changes at $z\sim2$.

Mergers and active galactic nuclei (AGN) feedback are some of the many processes which can contribute to quenching. Mergers may play a crucial role in quenching by suppressing cooling, inducing stellar feedback and/or AGN feedback, and triggering merger-induced shocks \citep{Hopkins08}. Gas-rich mergers may induce concentrated bursts of star formation before quenching occurs \citep[e.g.][]{Mihos1996, Hopkins2008, Wellons2015}. Simulations have shown that AGNs may quench galaxies via feedback mechanisms, although direct evidence is lacking and their dependence on redshift is still uncertain \citep{Springel05, Khalatyan08, Kaviraj11}. AGN feedback is thought to be the dominant feedback source for massive galaxies ($M_* > 10^{10} M_\odot$) and is expected to be more efficient as the mass of the galaxy increases, thus also increasing the efficiency of quenching \citep{Kaviraj07}.

Post-starburst galaxies (PSBs) provide an ideal laboratory for studying quenching, as they have undergone a rapid decline in their star formation in the past 1 Gyr. They can be observationally identified by an absence of star-formation-driven nebular emission lines (suggesting low current star formation) coupled with strong Balmer absorption (due to a stellar population dominated by A/F-type stars, indicating that a large fraction of the stellar mass was formed within the last Gyr) \citep[e.g.][]{French2021}. These features result in post-starburst galaxies having distinct spectral energy distribution (SED) shapes, which can be reliably identified using a principal component analysis (PCA) of broadband photometry \citep{Wild14}, color-color techniques \citep[e.g.][]{Whitaker2012}, and spectroscopic selections \citep[e.g.][]{Wild2020}. PSBs are thought to be an especially important quenching pathway at high redshift, as nearly all quiescent galaxies at $z>3$ appear to be PSBs \citep[e.g.][]{DEugenio2020, French2021}.

It has been established that galaxies can quench over a range of timescales \citep[e.g.][]{Moutard2016, Wu2018, EstradaCarpenter2020, Tacchella2022}, suggesting the existence of multiple quenching pathways, with fast quenching potentially becoming increasingly important at higher redshifts \citep{Rowlands2018, Belli2019}. In order for galaxies to be quenched quickly, the gas supply must have been consumed, heated, or removed through processes such as starbursts, feedback, and/or environmental effects. The typical masses and morphologies of quenching galaxies can vary with redshift \citep[e.g.][]{Wild16, Maltby18}, suggesting that the dominant quenching processes change over cosmic time. At low redshift ($z<1$), recently and rapidly quenched galaxies tend to have lower stellar masses \citep{Wild16}, are typically found in dense environments \citep{Wilkinson21}, and are more disc-dominated \citep{Maltby18}. In contrast, at $z>1$, the recently and rapidly quenched population is dominated by massive, compact, spherical galaxies \citep{Wild16, Almaini2017}, suggesting internal quenching following a gas-rich compaction event, such as a gas-rich merger. However, the dominant quenching mechanisms at different redshifts are still unresolved.

Prior to \textit{JWST}, we had been limited by small samples of high-resolution, rest-frame optical spectra of post-starburst and quiescent galaxies at higher redshifts ($z > 1$). We now have much larger samples from \textit{JWST} NIRSpec, which has helped constrain the star formation histories (SFHs) of early quiescent galaxies \citep[e.g.][]{Carnall23a, Nanayakkara2024, Park2024, Setton2024, Slob2024, Carnall2024}, and can provide insight into the relative importance of fast ($\sim 100$ Myr) and slow ($\sim 1$ Gyr) quenching processes. Additionally, many studies have found that a large fraction of quiescent galaxies at $z>2$ shows evidence for AGN activity based on line ratio diagnostics \citep[e.g.][]{Belli2024, DEugenio2024, Bugiani2024}, suggesting that AGN may play a role in quenching star formation, as discussed above. 

Outflows of neutral gas, likely driven by AGN, are prevalent and have been proposed as a dominant quenching mechanism for massive galaxies in the early Universe \citep{Davies2024, Wu2025, Valentino2025}. However, the role of AGN in quenching galaxies in the early Universe is still uncertain.\citep{Bevacqua2025}. 
There is some evidence that AGNs in local post-starburst galaxies may have been fueled by the same processes that caused the quenching, rather than the quenching being driven by AGN feedback \citep{Lanz2022}. Furthermore, \citet{Almaini2025} found no evidence for excess AGN activity in massive PSBs in the redshift range $1<z<3$. Instead, low average X-ray luminosities were found, consistent with weak AGN activity that traces the star formation rate.

To better understand the dominant quenching mechanisms at early epochs, we analyze a sample of 24 post-starburst and quiescent galaxies at redshifts $1<z<3$ from the JWST Early eXtragalactic Continuum and Emission Line Survey \citep[EXCELS;][]{Carnall2024}. The outline of this paper is as follows. In Section~\ref{sec:obs}, we describe our observations and methods. In Section~\ref{sec:res}, we discuss the physical properties and star formation histories of these galaxies, derived from full spectral fitting. We also explore the emission line properties of the sample. We discuss these results and propose possible quenching mechanisms in Section~\ref{sec:dis}, and present our summary in Section~\ref{sec:conc}. We assume a \cite{Kroupa2001} stellar initial mass function and a flat $\Lambda$CDM cosmology where necessary with $\Omega_M = 0.3$, $\Omega_ \Lambda = 0.7$, and $h=0.7$.

\section{Observations and Methods}
\label{sec:obs}

\subsection{Photometry}
We make use of 11 deep photometric bands from \textit{JWST}/PRIMER imaging (GO 1837; PI: Dunlop) and \textit{HST} imaging from CANDELS \citep{Grogin2011, Koekemoer2011} for all of the galaxies in our sample (\textit{JWST}/NIRCam filters F090W, F115W, F150W, F200W, F277W, F356W, F410M, and F444W;
\textit{HST} ACS filters F435W, F606W, and F814W). The photometric catalog will be presented in Maltby et al. (in prep), but in brief, the NIRCam F200W filter is used as the detection image. {We use the F200W filter, as opposed to a redder band, as it allows for better comparisons to ground-based studies in this field, which use the K-band filter as the detection image at $0.5<z<3$ \citep[e.g.][]{Almaini2017}.} Fluxes are measured in 0.5 arcsec-diameter apertures using \textsc{SourceExtractor} \citep{Bertin1996}. The photometry is corrected to the total flux with an aperture correction dependent on the PSF for each filter and using the F200W \texttt{MAGBEST} parameter from \textsc{SourceExtractor} (the total magnitude) applied to all the filters. We note that this is a different photometric catalog from the one used to select the original EXCELS sample. The original EXCELS photometric catalog uses the F356W band to select objects, uses 0.35 arcsec-diameter apertures, and uses the F356W FLUX\_AUTO parameter from \textsc{SourceExtractor} to correct to the total flux. However, the results presented in this paper remain the same regardless of which catalog is used. 

\subsection{Spectroscopy}

The galaxy sample presented in this work is selected from the \textit{JWST} EXCELS spectroscopic survey \citep[GO 3543; PIs: Carnall, Cullen; ][]{Carnall2024}.
The EXCELS sample selection is described in \cite{Carnall2024} and is briefly summarized below. The EXCELS survey is designed to obtain medium-resolution ($\text{R}\sim1000$) spectroscopy of galaxies in the \textit{JWST} Public Release IMaging for Extragalactic
Research (PRIMER) Ultra-Deep Survey (UDS) field. The 4 NIRSpec/MSA pointings include observations of massive quiescent galaxies selected from both the VANDELS survey UDS-HST selection catalog \citep{McLure2018} and from PRIMER UDS imaging \citep[GO 1837; PI: Dunlop; ][]{Dunlop2021}. Each pointing was observed for approximately 4 hours with the G140M and G395M gratings, and about 5.5 hours with the G235M grating, using the NRSIRS2 readout pattern.

To reduce the data, we process the raw level 1 products from the Mikulski Archive for Space Telescopes using v1.17.1 of the \textit{JWST} reduction pipeline\footnote{https://github.com/spacetelescope/jwst} and the Calibration Reference Data System (CRDS) context \texttt{jwst\_1322.pmap}. First, we use the default level 1 configuration, but turn on the advanced snowball rejection option. We also run the \texttt{clean\_flicker\_noise} step to account for the 1/f detector readout noise. We then run the default level 2 pipeline without the sky subtraction step. Residual artifacts in the 2D calibrated spectra from each of the three exposures, taken at different nod positions, are manually flagged and masked before performing sky subtraction using custom code. Finally, we run the level 3 pipeline to obtain the combined 2D flux calibrated spectra. We use a custom optimal extraction of the 2D spectra \citep{Horne1986} to obtain the 1D spectra featured in this work (see \citealp{Carnall2024}).

A flux calibration must be applied to the spectra to account for slit losses and spectrophotometric calibration issues. {There is an aperture mismatch between the NIRSpec MSA slits and the total-flux aperture photometry. By scaling the spectra to the total-flux photometry, we are assuming that the flux from the region covered by the slit is representative of the flux from the entire galaxy. We believe this is a reasonable assumption as PSBs at $z>1$ do not tend to have stellar-age gradients \citep{Maltby18}.} We integrate the spectra through the filters that overlap with the spectral data and scale the spectra to the fluxes by applying the mean normalization factor derived from all photometric points. As we often only have two photometric filters overlapping with the available spectra, we do not account for wavelength-dependent slit losses at this stage. These effects are instead addressed during our full spectral fitting analysis, where we introduce parameters to model them, as described in Section~\ref{subsec:pipes_fitting}. In the cases where multiple gratings are available, following \cite{Carnall2024}, we use {\sc spectres} \citep{Carnall2017} to resample the spectra of the shorter wavelength grating (which has a higher resolution, as the resolving power R is a function of wavelength for a given grating) onto the wavelength of the longer wavelength spectra in the overlap region. 
We then scale the G140M spectra to the mean flux of the region overlapping with the G235M spectra, and compute the mean pixel values from both gratings. A compilation of all the spectra is shown in Appendix~\ref{appendix_a}.

\subsection{Sample selection}
\label{sec:samp_sc}
From the EXCELS sample we select 24 galaxies within the redshift range $1<z<3$ that were identified as quiescent based on the VANDELS survey and PRIMER imaging (see Section~3 of \citealt{Carnall2024}). The VANDELS quiescent samples were selected based on the $z>1$ rest-frame UVJ color criteria proposed by \citet{Williams2009}. The quiescent sample from the PRIMER imaging was selected using a specific star-formation rate (sSFR) cut of $\text{sSFR} < 0.2/t_{\text{universe}}$. These two sample selection methods have been shown to be highly consistent \citep[e.g.][Stevenson et al. in prep]{Carnall2018}.
Additionally, there is one galaxy included in the survey that was selected as a PSB via a super-color analysis (described below) that was not in either the VANDELS or PRIMER passive samples. However, we find that the primary results presented in this paper do not depend on the selection method used.
We ensure that all of these galaxies have secure spectroscopic redshift measurements. We exclude one galaxy for which we were unable to get a reliable fit to the spectroscopic data (see Section~\ref{subsec:pipes_fitting}). The typical signal-to-noise ratio (S/N) per resolution element is 29 (ranging from 9 to 94). However, the S/N is slightly lower blueward of the Balmer/4000 Å break (with a median of 9, and ranging from 2 to 27). The galaxy with the lowest S/N is EXCELS–114929 is dusty. However, we include it in this sample, as the fit from \textsc{Bagpipes} shows it is quenched (see Sections~\ref{subsec:pipes_fitting} and~\ref{subsec:sfhs}).
The majority of our sample is observed only with the G140M grating (18 galaxies), with three being observed only with G235M and three observed with both gratings. The selection criteria and gratings used to select each source are listed in Table~\ref{tab1}.

\begin{table*}
\centering
\begin{tabular}{lllllllll}
\hline
 EXCELS ID   & $z$    & Grating     & $\log M^*/M_\odot$      & Selection              & SC class   & Burst age (Gyr)        & Lick H$\delta_A$ ($\text{\AA}$)   & EW [OII] ($\text{\AA}$)   \\
\hline
 $101499$    & $1.09$ & G140M       & $10.25^{+0.05}_{-0.06}$ & $\textnormal{VANDELS}$ & SF         & $1.51^{+0.3}_{-0.5}$   & $7.44^{+0.4}_{-0.4}$              & $8.18^{+0.4}_{-0.4}$      \\
 $114127$    & $2.53$ & G235M       & $10.56^{+0.05}_{-0.06}$ & $\textnormal{PRIMER}$  & Q          & $1.30^{+0.4}_{-0.3}$   & --                                & --                        \\
 $114633$    & $2.61$ & G140M       & $10.54^{+0.06}_{-0.05}$ & $\textnormal{PRIMER}$  & PSB        & $0.69^{+0.1}_{-0.09}$  & $6.47^{+0.5}_{-0.5}$              & $2.71^{+2}_{-0.5}$        \\
 $114929$    & $2.18$ & G140M       & $11.18^{+0.03}_{-0.04}$ & $\textnormal{PRIMER}$  & SF (dusty) & $1.64^{+0.3}_{-0.5}$   & $8.45^{+1}_{-1}$                  & $6.71^{+2}_{-2}$          \\
 $122839$    & $1.41$ & G140M       & $10.87^{+0.04}_{-0.03}$ & $\textnormal{VANDELS}$ & SF         & $1.68^{+0.2}_{-0.5}$   & --                                & --                        \\
 $123852$    & $2.52$ & G140M       & $10.82^{+0.04}_{-0.05}$ & $\textnormal{PRIMER}$  & SF         & $1.65^{+0.2}_{-0.4}$   & --                                & $27.68^{+1}_{-1}$         \\
 $127460$    & $2.53$ & G140M/G235M & $10.85^{+0.03}_{-0.03}$ & $\textnormal{PRIMER}$  & PSB        & $0.52^{+0.1}_{-0.1}$   & $7.05^{+0.3}_{-0.3}$              & $6.43^{+0.3}_{-0.3}$      \\
 $43944$     & $2.30$ & G235M       & $10.51^{+0.06}_{-0.04}$ & $\textnormal{PRIMER}$  & PSB        & $1.28^{+0.4}_{-0.2}$   & --                                & --                        \\
 $54835$     & $2.48$ & G140M       & $10.78^{+0.03}_{-0.03}$ & $\textnormal{PRIMER}$  & Q          & $0.51^{+0.1}_{-0.2}$   & $10.48^{+0.4}_{-0.3}$             & $11.13^{+1}_{-2}$         \\
 $55414$     & $1.47$ & G140M       & $11.04^{+0.05}_{-0.05}$ & $\textnormal{VANDELS}$ & Q          & $1.57^{+0.3}_{-0.2}$   & $0.76^{+0.2}_{-0.3}$              & $0.96^{+0.2}_{-0.2}$      \\
 $60232$     & $1.09$ & G140M       & $10.19^{+0.05}_{-0.04}$ & $\textnormal{VANDELS}$ & Q          & $1.82^{+0.1}_{-0.2}$   & --                                & --                        \\
 $63089$     & $2.09$ & G140M/G235M & $10.55^{+0.06}_{-0.05}$ & $\textnormal{PRIMER}$  & Q          & $0.68^{+0.2}_{-0.1}$   & $4.12^{+0.4}_{-0.3}$              & $0.00^{+0}_{-0}$          \\
 $63832$     & $2.32$ & G140M       & $10.86^{+0.04}_{-0.03}$ & $\textnormal{PRIMER}$  & PSB        & $0.83^{+0.1}_{-0.1}$   & $4.32^{+0.4}_{-0.3}$              & $2.05^{+2}_{-0.9}$        \\
 $65234$     & $2.23$ & G140M/G235M & $10.18^{+0.03}_{-0.03}$ & $\textnormal{PRIMER}$  & PSB        & $0.45^{+0.08}_{-0.07}$ & $9.34^{+0.3}_{-0.3}$              & --                        \\
 $69069$     & $2.15$ & G140M       & $10.33^{+0.04}_{-0.04}$ & $\textnormal{PRIMER}$  & PSB        & $0.53^{+0.1}_{-0.1}$   & $7.75^{+0.4}_{-0.4}$              & $0.00^{+0}_{-0}$          \\
 $69759$     & $1.53$ & G140M       & $11.36^{+0.03}_{-0.03}$ & $\textnormal{VANDELS}$ & Q          & $1.67^{+0.2}_{-0.3}$   & $-0.84^{+0.5}_{-0.4}$             & --                        \\
 $72186$     & $2.32$ & G140M       & $10.96^{+0.02}_{-0.02}$ & $\textnormal{PRIMER}$  & PSB        & $1.01^{+0.08}_{-0.08}$ & $6.02^{+0.2}_{-0.2}$              & $1.78^{+0.2}_{-0.2}$      \\
 $90549$     & $1.09$ & G140M       & $10.45^{+0.06}_{-0.07}$ & $\textnormal{VANDELS}$ & Q          & $1.76^{+0.2}_{-0.3}$   & --                                & --                        \\
 $90591$     & $1.09$ & G140M       & $10.15^{+0.05}_{-0.05}$ & $\textnormal{VANDELS}$ & PSB        & $0.84^{+0.2}_{-0.1}$   & --                                & --                        \\
 $91726$     & $1.10$ & G140M       & $10.85^{+0.03}_{-0.04}$ & $\textnormal{VANDELS}$ & Q          & $1.58^{+0.3}_{-0.2}$   & $0.91^{+0.3}_{-0.2}$              & $10.47^{+0.3}_{-0.2}$     \\
 $93227$     & $1.09$ & G140M       & $10.45^{+0.05}_{-0.06}$ & $\textnormal{VANDELS}$ & Q          & $1.26^{+0.2}_{-0.2}$   & $1.74^{+0.2}_{-0.2}$              & $2.82^{+0.1}_{-0.2}$      \\
 $94982$     & $1.83$ & G140M       & $10.97^{+0.03}_{-0.02}$ & $\textnormal{SC PSBs}$ & PSB        & $0.84^{+0.05}_{-0.07}$ & $6.30^{+0.1}_{-0.1}$              & $2.26^{+0.1}_{-0.1}$      \\
 $97765$     & $2.29$ & G140M       & $10.45^{+0.03}_{-0.03}$ & $\textnormal{PRIMER}$  & PSB        & $0.79^{+0.08}_{-0.1}$  & $3.84^{+0.3}_{-0.3}$              & $1.69^{+0.7}_{-0.9}$      \\
 $98447$     & $2.81$ & G235M       & $11.34^{+0.02}_{-0.02}$ & $\textnormal{PRIMER}$  & PSB        & $0.28^{+0.09}_{-0.08}$ & --                                & --                        \\
\hline
\end{tabular}
\caption{Measured properties of our sample of galaxies. The stellar masses and burst ages come from \textsc{Bagpipes}. The selection indicates how the galaxy was chosen, as described in Section~\ref{sec:samp_sc}. The SC class denotes the super-color classification based on photometry (see Section~\ref{sec:samp_sc}), where galaxies are classified as post-starburst (PSB), quiescent (Q), or star-forming (SF). We define H$\delta_A$ as positive for absorption, while the equivalent width of [OII] is defined as positive in emission.}
\label{tab1}
\end{table*}

We further subdivide our sample by calculating ``super-colors" following the methodology of \cite{Wild14}, which has been shown to cleanly separate post-starburst galaxies from the rest of the galaxy population when applied to broadband photometric data. This technique involves identifying three “super-colors” (or principal component amplitudes) that represent a wide variety of shapes: the first super-color determines how red the SED is, the second super-color determines the strength of the Balmer or 4000Å break, and the third super-color determines the shape around the Balmer or 4000Å break. The super-colors are a linear combination of the available photometric bands from \textit{HST} and \textit{JWST} that maximize the variance of model SEDs using a PCA. Super-colors are thus similar to traditional colors without forcing the data to fit model SEDs. Within ``super-color space" boundaries can be drawn to separate post-starburst, quiescent, star-forming, and dusty star-forming galaxies. These boundaries have been spectroscopically verified at $z<1.4$ \citep{Maltby2016, Wilkinson21}. The super-color classifications for the PRIMER field will be presented in de Lisle et al. (in prep). Note that a different eigenbasis is used, as we are using \textit{HST} and \textit{JWST} filters, so the boundaries differ from those in \cite{Wild14}. 

\begin{figure}
    \centering	
    \includegraphics[width=\linewidth]{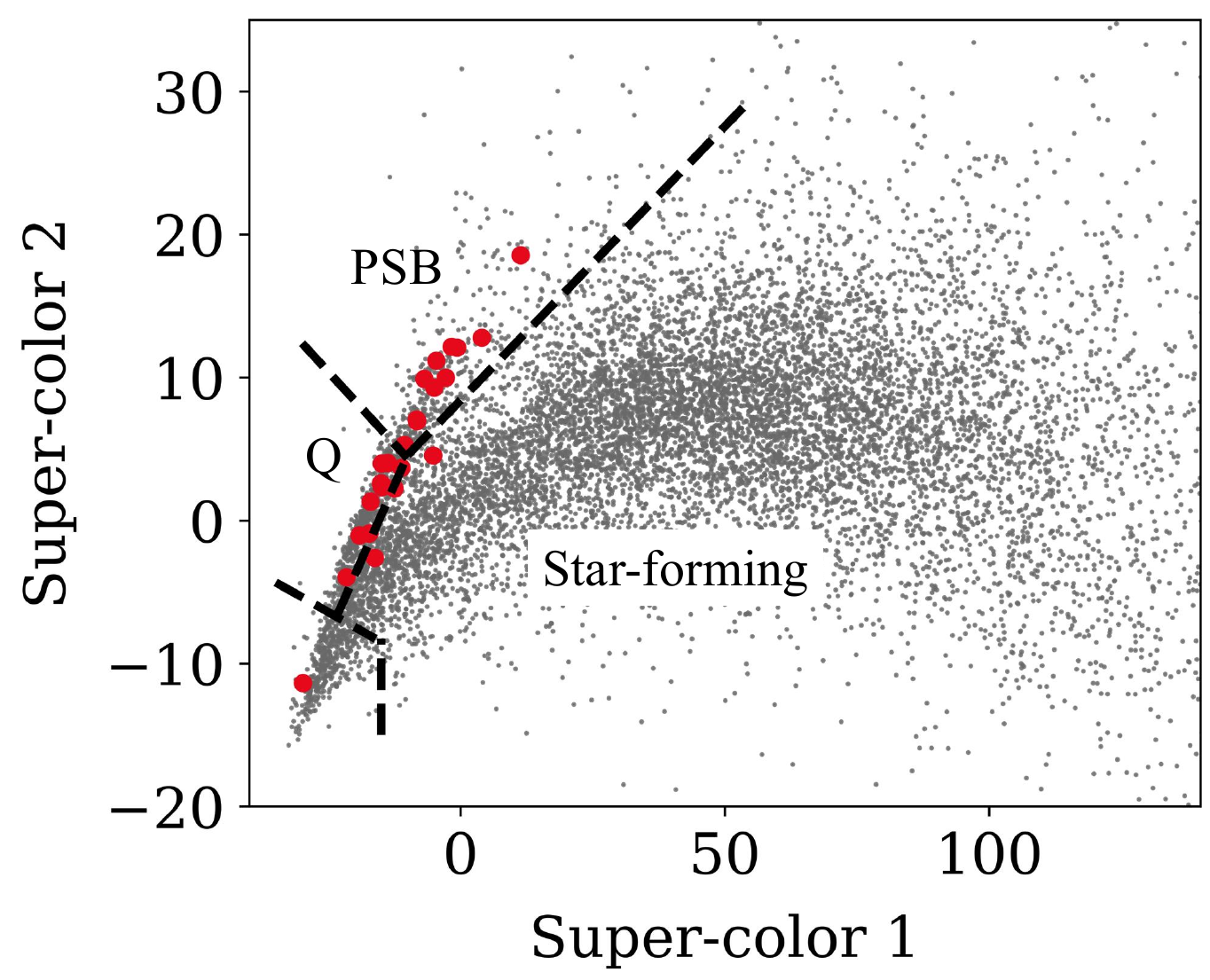}
    \caption{Super-color 1 (set by how red or blue the galaxy SED is) versus super-color 2 (determined by the strength of the 4000\AA\, or Balmer break) based on photometry. Our sample of spectroscopically confirmed post-starburst and quiescent galaxies is shown in red and the underlying sample from the entire PRIMER field (de Lisle et al. in prep) is shown in gray. The dashed black lines indicate the boundaries between the post-starburst, quiescent, and star-forming regions.}
    \label{Fig::pca_confirm}
\end{figure}

The super-color classifications for the galaxies in our sample are listed in Table~\ref{tab1} and shown in Figure~\ref{Fig::pca_confirm}. Twenty of our galaxies fall within the quiescent or post-starburst regions of the diagram, with three galaxies classified as star-forming lying close to the quiescent boundaries and one galaxy classified as a dusty star-forming galaxy. There are 11 galaxies that fall within the PSB region of the super-color diagram and 9 galaxies classified as quiescent. Throughout the paper, we use the term `PSB' to refer to super-color-selected PSBs.

\subsection{Spectral Fitting} 
\label{subsec:pipes_fitting}
We use the Bayesian spectral fitting code \textsc{Bagpipes} \citep{Carnall2018, Carnall2019} to jointly fit our spectroscopic data and all available \textit{HST} ACS and \textit{JWST}/NIRCam photometry. Additionally, we make use of archival spectra from VANDELS \citep{McLure2018, Pentericci2018, Garilli2021}, observed with the VIMOS instrument. There are four galaxies in our sample (ID-101499, ID-55414, ID-91726, ID-93227) with VANDELS spectra that we fit in conjunction with the NIRSpec spectra. We bin the VANDELS spectra by 2 pixels and fit the wavelength range $5200\textrm{Å}$ to $9250 \textrm{Å}$, following \citet{Carnall2019}. We note that the results remain consistent when not including the VANDELS spectra. 

Our priors (summarized in Table~\ref{tab:pipes_priors}) are similar to those described in \cite{Carnall2024}, apart from the SFH and dust model. We use the 2016 updated version of the \cite{Bruzual2003} spectral synthesis models, which uses the MILES stellar library \citep{Sanchez2006, FB2011}. This library spans $3540-7350 \textrm{\AA}$, so we only fit our spectroscopic data within this range in rest-frame wavelengths. We mask regions of the spectra that may contain emission lines with a $\pm 5 \textrm{Å}$ mask in the rest-frame. We include a nebular component with the ionization parameter fixed to $\log U = -3$. As done in \cite{Carnall2024}, we include a 38th order Chebyshev polynomial to account for spectrophotometric calibration issues and slit losses. This is about 1 order per $100\textrm{Å}$ of rest-frame wavelength coverage, following the approach of \citet{Conroy2012} and \citet{Conroy2014}. The priors on the redshifts are from the spectroscopic redshifts described in \cite{Carnall2024}. To address the possibility of underestimating the observed uncertainties, we introduce a multiplicative factor on the error of the spectroscopic data, allowing it to vary from $1-10$. This high-order polynomial has the potential to absorb continuum features. To investigate the impact of this Chebyshev polynomial, we ran Bagpipes with a low- and high- order polynomial (10 and 76, respectively), following \citet{Carnall2024}. We find that using a higher- and lower-order polynomial does not significantly impact our results. Specifically, we have recreated all the figures presented in this paper and calculated the quenching times, visibility times, and SFHs, and find that none of our conclusions are changed. The changes to the calibration component as a function of wavelength typically vary about $\pm 5$ percent. The median of the posterior distribution of the error-scaling parameter $a$ ranges from 1.4 to 2.5. We use the nested sampling algorithm \textsc{MultiNest} \citep{Feroz2019} to sample the model posterior, which is accessed with the \textsc{PyMultiNest} package \citep{Buchner2014}. We assume a \cite{Kroupa2001} stellar initial mass function. We adopt the two-component dust attenuation model described in \cite{Wild2007}, which assumes a fixed power-law exponent of $n=0.7$ for the wavelength dependence of dust attenuation in the interstellar medium (ISM). Stars that are younger than 10 Myr have a fixed slope of $n=1.3$ and are assumed to be more attenuated by a factor of $\eta$ (modeled with a Gaussian prior) due to the birth clouds surrounding the stars. We allow the metallicity to vary from 0.01 to 2.5 $Z_\odot$. The metallicity grid points of the stellar library interpolated by \textsc{Bagpipes} is (0.005, 0.02, 0.2, 0.4, 1., 2.5) $Z_\odot$.

\begin{table*}
    \centering
    \begin{tabular}{ccccc}
    \hline
        Model Component & Parameter & Range & Prior \\ \hline
        SFH & $\log_{10}(M_*/M_\odot)$ & (9,13) & Uniform\\
        & $t_{\text{form}}$ / Gyr & (0.4, 8) & Uniform\\
        & $\tau_e$ / Gyr & (0.3, 10) & Uniform\\
        & $t_{\text{burst}}$ / Gyr & (0,2) & Uniform\\
        & $f_{\text{burst}}$  & (0, 1) & Uniform\\
        & $\alpha$  & (0.01, 1000) & Logarithmic\\
        & $\beta$  & 250 & --\\ \hline
        Metallicity & $Z$ / $Z_\odot$ & (0.01, 2.5) & Logarithmic \\ \hline
        Dust & $A_V$ / mag & (0,2) & Uniform \\
        & Birthcloud factor $\eta$  &(1,4) & Gaussian ($\mu=3$, $\sigma = 1$)\\ \hline
        Noise & White noise scaling $a$ & (1,10) & Logarithmic\\ \hline
        Miscellaneous & Redshift & ($z_{spec} \pm 0.01$) & Gaussian\\
         &  $\sigma$/km s$^{-1}$ & (40, 500) & Logarithmic\\
    \hline
    \end{tabular}
    \caption{Model priors used for \textsc{Bagpipes} fitting, as described in Section~\ref{subsec:pipes_fitting}. All logarithmic priors are defined in base 10.}
    \label{tab:pipes_priors}
\end{table*}

\begin{figure}
    \centering	\includegraphics[width=\linewidth]{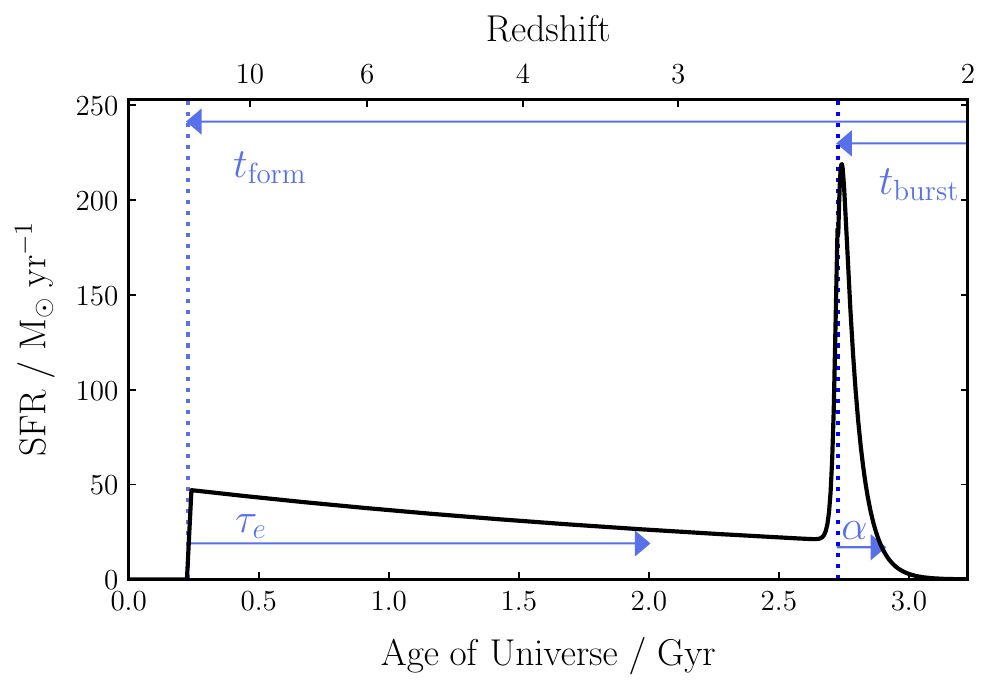}
    \caption{An example SFH (see equation~\ref{eq::sfh}) used to fit our sample of galaxies. The galaxy in this model has a redshift of $z=2$, a burst age of $t_{\text{burst}} = 0.5$ Gyr, a formation age of $t_{\text{form}} = 3$ Gyr, and has formed $10^{11} M_\odot$ of stars. It has a burst mass fraction of 20 percent ($f_{\text{burst}} = 0.2$), an exponential decay timescale of $\tau_{\text{e}} = 3$ Gyr, and $\alpha = 50$ describing the declining timescale of the burst. }
    \label{Fig::sfh_ex}
\end{figure}

We assume the two-component parametric SFHs described in \cite{Wild2020}. The star formation rate as a function of time is given by
\begin{equation}
\begin{aligned}
    \psi(t) \propto &\frac{1-f_{\text{burst}}}{\int \psi_e \text{d}t} \times \psi_e(t) \bigg|_{t_{\text{form}} > t > t_{\text{burst}}} \\
    &+ \frac{f_{\text{burst}}}{\int \psi_e \text{d}t} \times \psi_{\text{burst}}(t).
\end{aligned}
\label{eq::sfh}
\end{equation}
In addition to a double power-law component $\psi_{\text{burst}}(t)$, this SFH also includes an exponential decay component $\psi_e(t)$ to account for older stellar populations. The parameter $f_{\text{burst}}$ determines the fraction of mass formed during the starburst, and $t_{\text{form}}$ denotes the lookback time when the older component started to form (which is limited to the age of the Universe when the age of the Universe is less than the prior's upper bound of 8 Gyr). The burst age $t_{\text{burst}}$ is defined as the time since the peak of the starburst. The two components are described by
\begin{equation}
\begin{aligned}
    \psi_e(t) = \exp^{\frac{-t}{\tau_{\text{e}}}}
\end{aligned}
\end{equation}
and
\begin{equation}
\begin{aligned}
    \psi_{\text{burst}}(t) = \left[ \left(\frac{t}{t_{burst}} \right)^{\alpha} + \left(\frac{t}{t_{burst}} \right)^{-\beta} \right]^{-1}.
\end{aligned}
\end{equation}
The exponential decay timescale of the older population is denoted as $\tau_{\text{e}}$, and $\alpha$ and $\beta$ set the declining and rising slopes of the starburst. We set $\beta = 250$, which is a typical value found for younger starbursts, as the variable $\beta$ was found to be poorly constrained and had no impact on the results \citep{Wild2020, Leung2024}. We show an example SFH in Figure~\ref{Fig::sfh_ex}.

\begin{figure*} 
    \centering
    \includegraphics[width=0.95\textwidth]{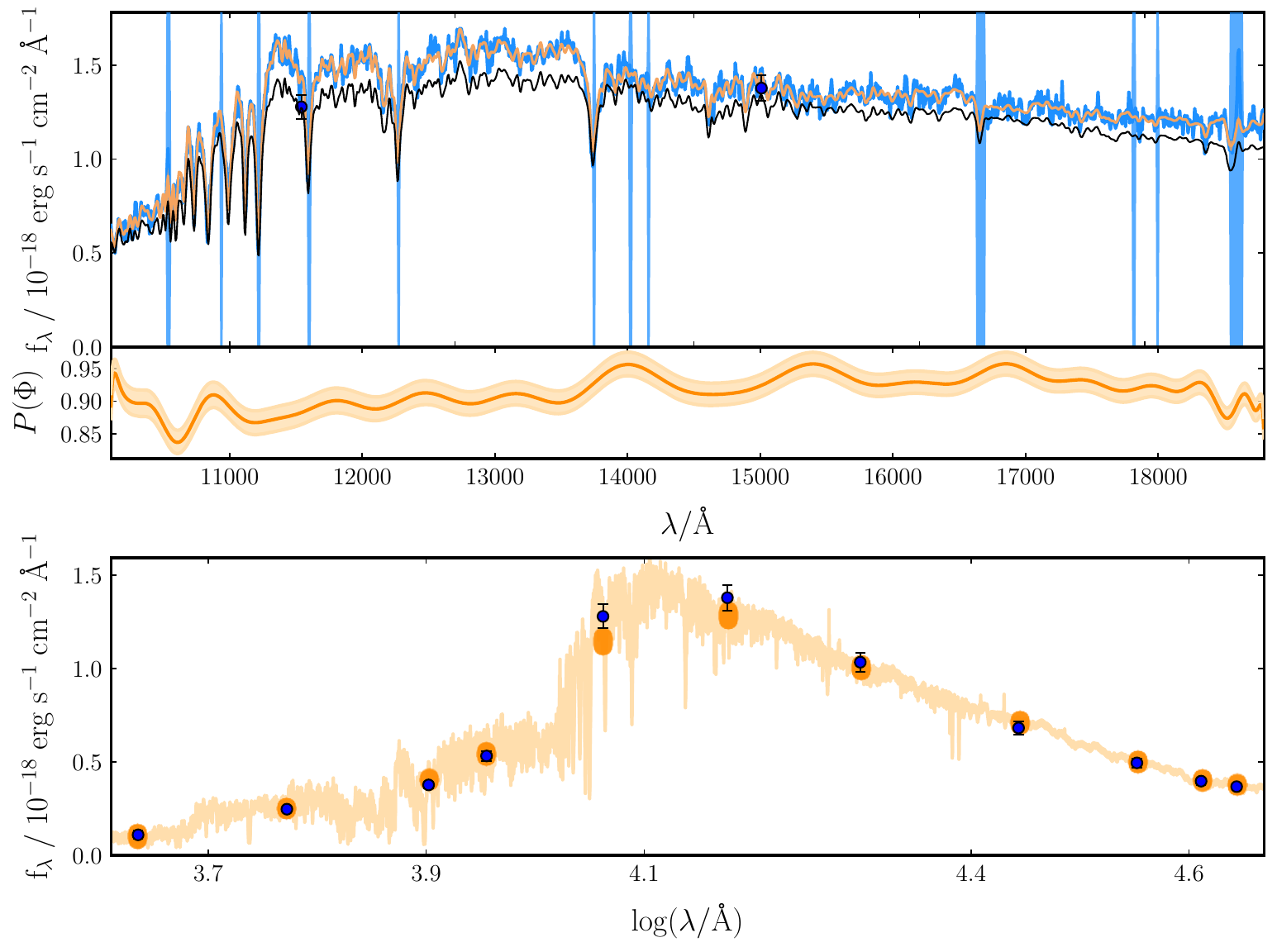}
    \caption{Example \textsc{Bagpipes} fit of a post-starburst galaxy at $z=1.83$ (EXCELS -- 94982). \textit{Top:} The input spectrum is shown in blue and the \textbf{median of the} posterior fit is shown in orange. The photometric measurements are shown as blue points and the black line denotes the fitted model without calibration corrections. The blue shading indicates regions masked during fitting. The lower panel shows the spectrophotometric calibration component. \textit{Bottom:} The input photometry is shown as blue points, the posterior fitted photometry is shown in orange, and the orange line shows the posterior SED.}
    \label{Fig::pipes_fit}
\end{figure*}

The free parameters and their corresponding priors of our \textsc{Bagpipes} model are summarized in Table~\ref{tab:pipes_priors}. Errors on the properties derived from \textsc{Bagpipes} are calculated as the 16th to 84th percentile range. An example of one of the \textsc{Bagpipes} fits for a galaxy in our sample with a burst age of $0.84^{+0.05}_{-0.07}$ Gyr and a redshift of $z=1.83$ is shown in Figure~\ref{Fig::pipes_fit}. We note that there is a slight offset between the model and input spectrum, which is likely due to slit losses and any issues with the spectrophotometric calibration, that our calibration component corrects for.

\subsection{Spectral line fitting}

To obtain emission line fluxes, we use the Penalized PiXel-Fitting (pPXF) method \citep{Cappellari2023}, which is upgraded from \citet{Cappellari2017} and originally described in \citet{Cappellari2004}. We use templates from the Flexible Stellar Population Synthesis \citep[FSPS;][]{Conroy2009, Conroy2010} models.
We perform an initial fit with an additive Legendre polynomial of degree 10 to obtain the stellar velocity and velocity dispersion. If there are multiple gratings, we set the instrumental resolution of the overlapping region to be that of the higher wavelength grating. We then perform a fit optimized to the gas component, holding the stellar velocity and velocity dispersion constant. The gas is modeled as a single kinematic component. For this fit, we only include a multiplicative Legendre polynomial of degree 10. We correct for Balmer absorption from the stellar continuum fit, as we only measure fluxes from the gas emission line fits.

The Lick H$\delta_A$ index \citep{Worthey1997} is often used to select for post-starburst galaxies \citep[e.g.][]{Chen2019, French2021}. We measure the Lick H$\delta_A$ index for all galaxies with available wavelength coverage. Errors are estimated by adding random noise to each spectrum. This noise is uniformly distributed between -1 and 1 times the error spectra. We then recalculate H$\delta_A$ for each perturbed spectrum. We also calculate the equivalent widths (EWs) of [OII]$\lambda$3727 and H$\alpha$, which are often used to select against current star formation \citep[e.g.][]{Chen2019, French2021}. We calculate these equivalent widths from the gaussian fits from pPXF. Errors are similarly calculated by perturbing each spectra with noise that is uniformly distributed between -1 and 1 times the error spectra. We repeat the fitting procedure described above and calculate the errors as the 16th and 84th percentiles. 
We list these measurements for our sample in Table~\ref{tab1}.

\section{Results}
\label{sec:res}

\subsection{Star formation histories}
\label{subsec:sfhs}
\begin{figure}
    \centering	
    \includegraphics[width=\linewidth]{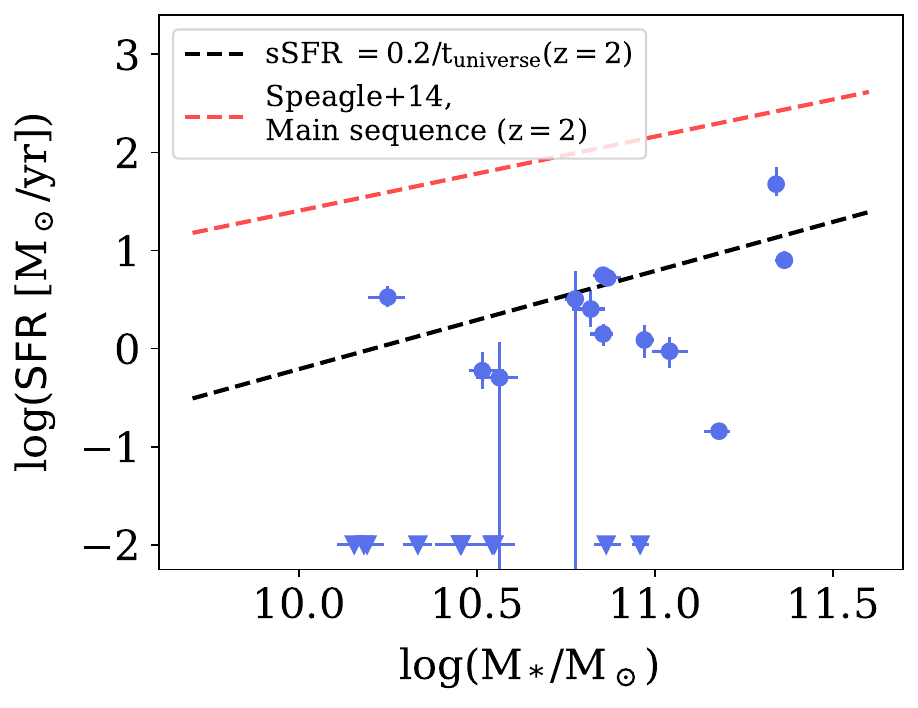}
    \caption{Star formation rate as a function of galaxy stellar mass, where both properties have been derived from our full spectral fitting analysis described in Section~\ref{subsec:pipes_fitting}. Triangles indicate values that fall below log(SFR)$<-2$. The dashed black line indicates where $\text{sSFR} < 0.2/t_{\text{universe}}$ at a redshift of $z=2$. The dashed red line indicates the star-forming main sequence at $z=2$ from \cite{Speagle2014}.}
    \label{Fig::sfr_mass}
\end{figure}

In Figure~\ref{Fig::sfr_mass} we show the star formation rate (SFR) averaged over the last 100 Myr derived from \textsc{Bagpipes} as a function of stellar mass for the 24 quiescent and post-starburst galaxies in our sample. All of the galaxies currently lie below the star-forming main sequence at $z=2$ from \cite{Speagle2014}, and the majority of the sample satisfies $\text{sSFR} < 0.2/t_{\text{universe}}$, a common criteria in the literature for selecting quiescent galaxies \citep[e.g.][]{Pacifici2016, Carnall2024}.

\begin{figure*}
    \centering	\includegraphics[width=0.95\linewidth]{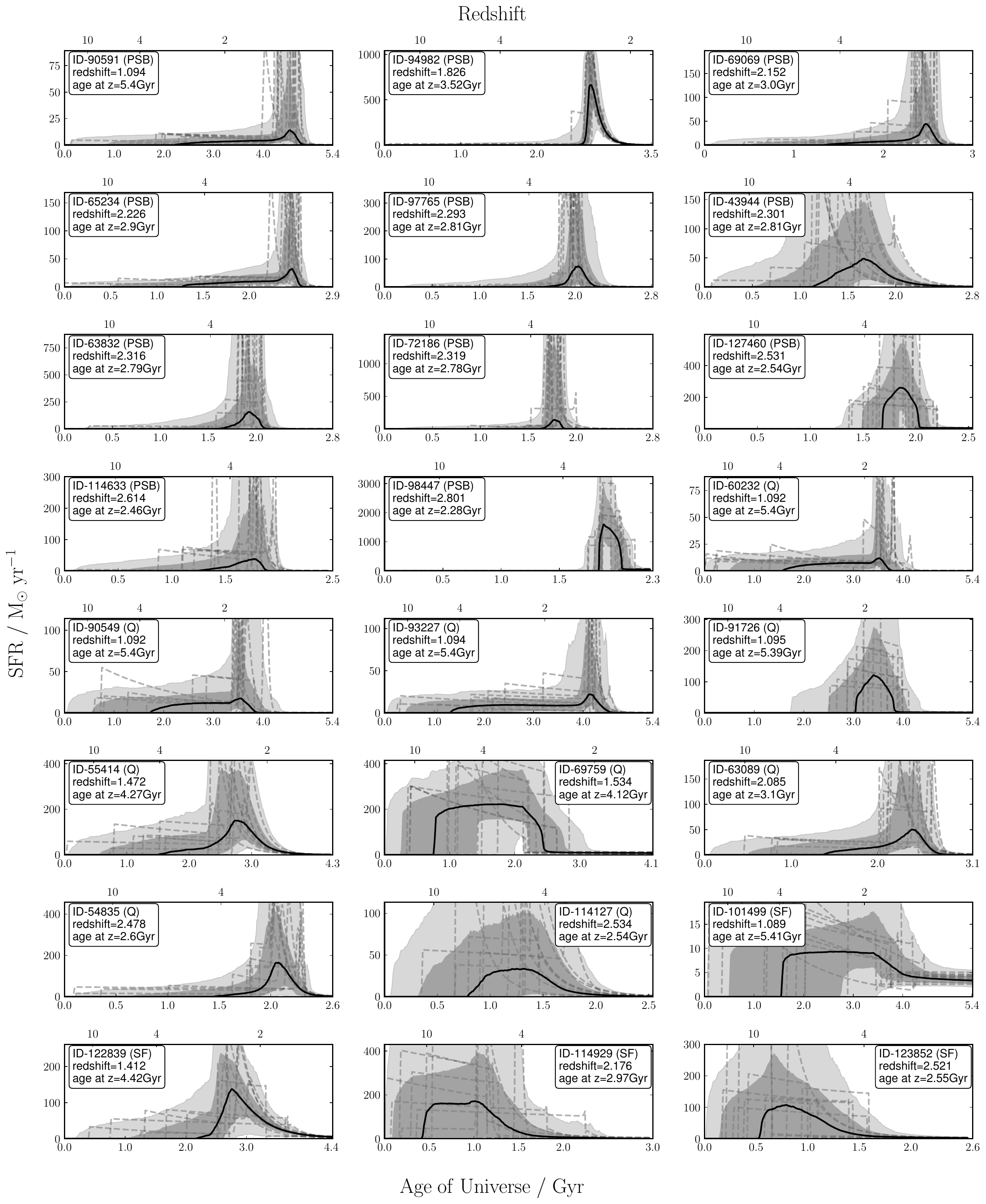}
    \caption{The star formation histories derived from \textsc{Bagpipes} for the galaxies in our sample. The labels on each plot indicate the EXCELS-ID followed by its super-color classification (Q, PSB, SF), as described in Section~\ref{sec:samp_sc}. The black line indicates the median posterior in each age bin, the gray shading indicates the 68 percent confidence interval, and the light gray shading indicates the 95 percent confidence interval of the SFHs. The gray dashed lines show 10 random samples from the posterior distributions.}
    \label{Fig::sfh}
\end{figure*}

We show a montage of the star formation histories (SFHs) fitted by \textsc{Bagpipes} for our sample of galaxies in Figure~\ref{Fig::sfh}. The majority of our sample had high peak star formation rates ($\sim 100-1000 \textrm{ M}_\odot \textrm{yr}^{-1}$), followed by rapid quenching and have low current star-formation rates. However, we observe a diverse range of SFHs. Some galaxies exhibit a well-defined burst of star formation followed by rapid quenching, while others appear to undergo a more prolonged period of star formation without a strong starburst. The star formation histories of $z\sim 1$ post-starburst galaxies presented in \cite{Wild2020} display a similar diversity to the SFHs found in this sample. 

To assess the robustness of our conclusions to the choice of the SFH prior, we compare the SFHs derived here with those obtained using a double power-law prior following \citet{Carnall2024}. The conclusions made throughout the paper remain the same when using a double power-law prior (the stellar mass, quenching timescales, and peak historical star formation rates are consistent within the errors). The exact duration of star formation for a few galaxies (e.g. EXCELS -- 69759 and EXCELS -- 114127) in our sample is poorly constrained due to large uncertainties in their SFHs, regardless of the prior.

\subsection{Super-colors}
\label{sec:res_SCs}
In this section, we discuss the photometric classifications of our sample. We will use these classifications, based on photometry alone, to compare differences between our sample and the $z\sim1$ galaxies in \cite{Wild2020}. Understanding how well these photometric classifications work and their limitations is important, given that many galaxies lack spectroscopic data.

The criterion used in \citet{Wild2020} to select PSBs is H$\delta_A>5 \text{\AA}$ and EW [OII]$<5 \text{\AA}$.
Of all the super-color selected PSBs in our sample with wavelength coverage of H$\delta_A$ and [OII], all but two satisfy H$\delta_A>5 \text{\AA}$ (but all satisfying H$\delta_A>3.8 \text{\AA}$) and all but one satisfy EW [OII]$<5 \text{\AA}$ (but all satisfying EW [OII]$<6.5 \text{\AA}$). Thus, the only galaxies in the PSB region that do not satisfy the criterion in used in \citet{Wild2020} are close to the limits. Of the six super-color selected quiescent galaxies with wavelength coverage of H$\delta_A$, one has H$\delta_A>5 \text{\AA}$. Three of the five quiescent galaxies with wavelength coverage of [OII] have an EW [OII]$<5 \text{\AA}$, with two having an EW [OII]$>10 \text{\AA}$. One of the quiescent galaxies with a high EW [OII] has H$\delta_A<1 \text{\AA}$.
However, the other quiescent galaxy (EXCELS -- 54835) has both a high  H$\delta_A$ and EW [OII], suggesting there is some contamination in the quiescent region based on traditional spectroscopic selection methods. 
This confirms for the first time that the super-color selected PSB and quiescent galaxies are robust above redshifts of $z\sim 1.4$ up to $z\sim 3$. Further investigation of the super-color selection method will be presented in de Lisle et al. (in prep), which will include a larger sample of ground-based spectra. 

\begin{figure}
    \centering	
    \includegraphics[width=\linewidth]{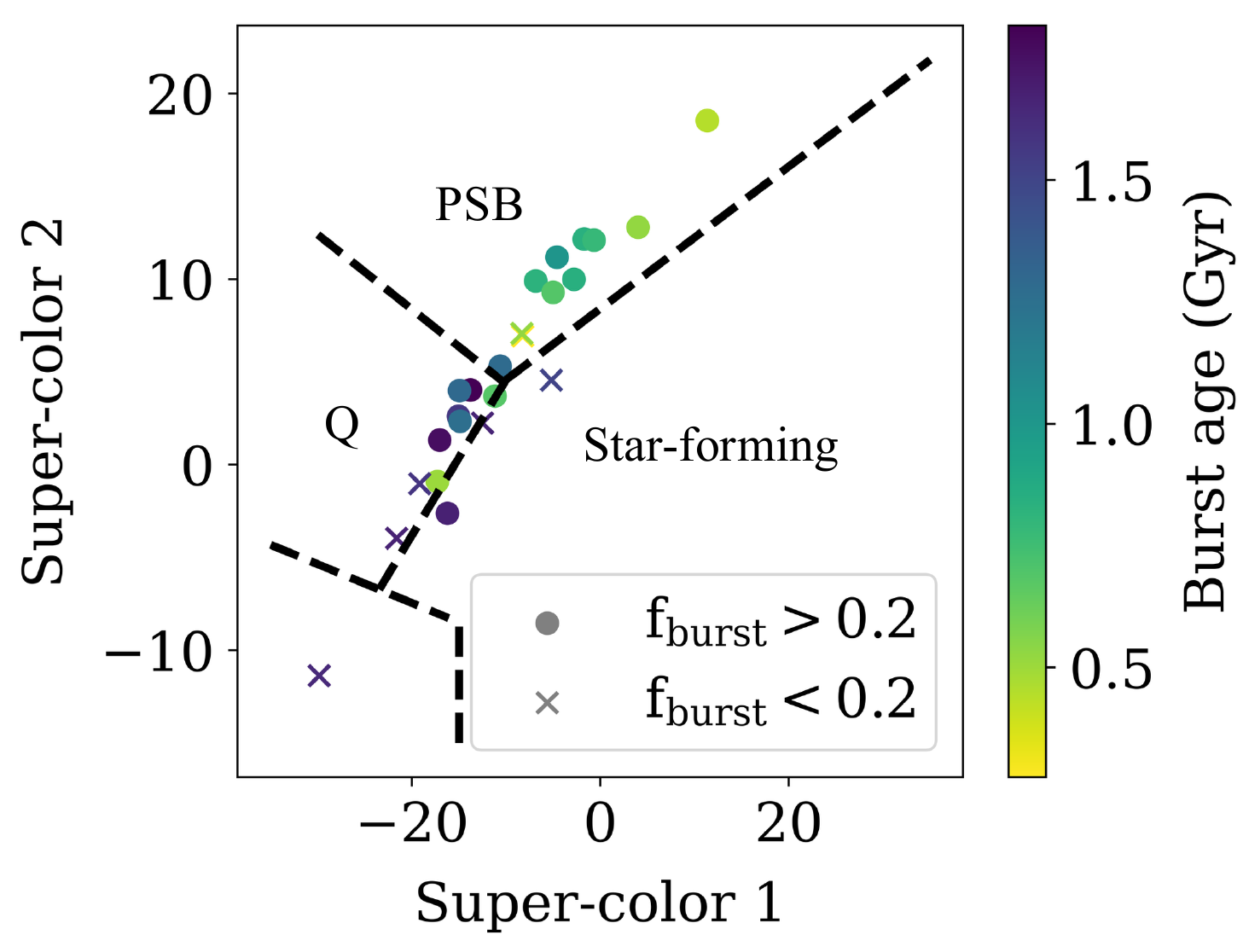}
    \caption{Super-color 1 versus super-color 2 based on photometry, as in Figure~\ref{Fig::pca_confirm}. The points are colored according to the burst age of the galaxy \textbf{(defined as the time since the peak of the starburst)} from \textsc{Bagpipes}. The size of the points indicates whether the burst mass fraction is greater or less than 0.2.}
    \label{Fig::pca_burstage}
\end{figure}

Figure~\ref{Fig::pca_burstage} shows the burst ages (the time since the peak of the starburst) of the galaxies in super-color space, as derived using \textsc{Bagpipes}. The galaxies in the post-starburst region of the diagram tend to have younger burst ages than the galaxies falling in the quiescent region, as expected. The median burst age of galaxies is 0.8 Gyr within the post-starburst region, and 1.5 Gyr within the quiescent region. Within the post-starburst region, there is a tentative trend in burst age, with younger burst ages having higher super-color 1 and super-color 2, although we are limited by a small number of post-starburst galaxies. Such a trend is not present in the quiescent region of the diagram. Within the post-starburst region, we find that galaxies closer to the quiescent region with young burst ages have lower burst mass fractions. This may be due to the positive correlation between the burst age and the burst mass fraction, possibly resulting from high uncertainties in the burst mass fraction or from effects related to the degeneracy between the burst age and the mass \citep{Leung2025}. There is a known degeneracy between the age and strength of the starburst, as younger, weaker bursts appear similar to older, stronger bursts. Therefore, scatter between the burst age and the position in super-color space is expected when relying on photometry alone \citep{Wild2007}. 

\begin{figure}
    \centering	
    \includegraphics[width=\linewidth]{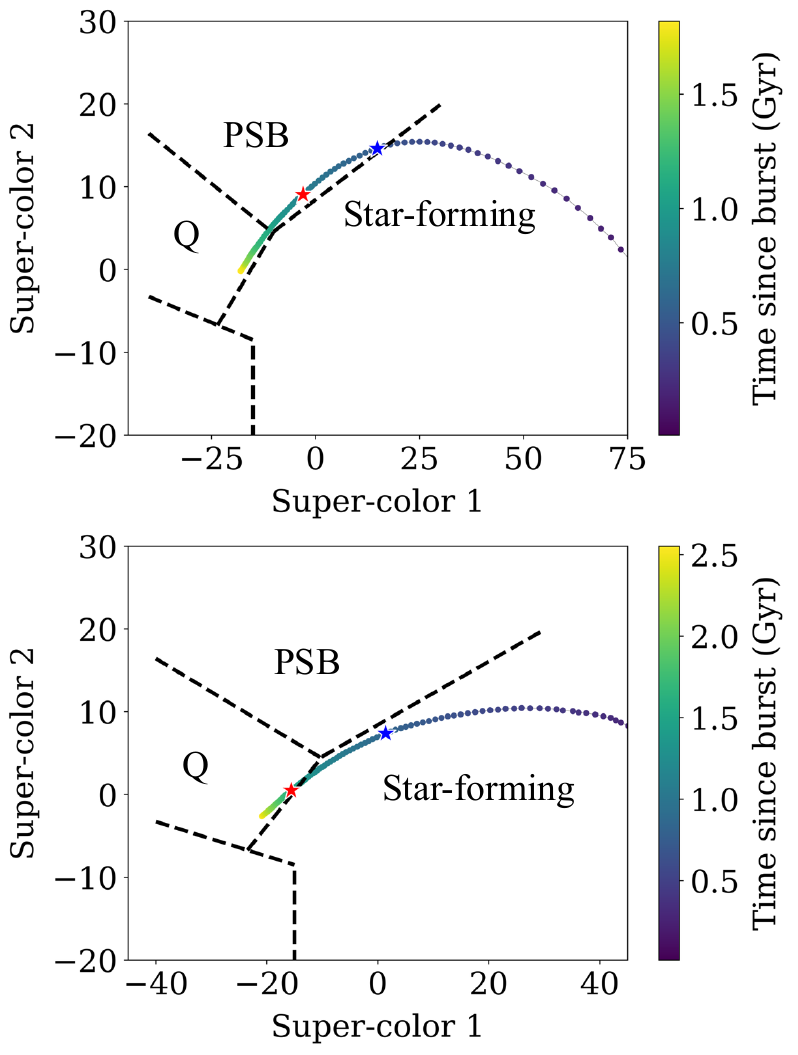}
    \caption{Change in super-colors over time based on the star-formation history from \textsc{Bagpipes} for EXCELS-94982 (top) and EXCELS-55414 (bottom). The red star indicates the current position of the galaxy in super-color space. The blue star indicates when the galaxy became quenched ($\text{sSFR} < 0.2/t_{\text{universe}}$). The dashed black lines indicate the boundaries between the post-starburst, quiescent, and star-forming regions, as in Figure~\ref{Fig::pca_burstage}.}
    \label{Fig::sc_track}
\end{figure}

Using the star formation histories derived from \textsc{Bagpipes}, we can predict how the galaxies in our sample move through super-color space. We generate mock photometry starting from the peak of the most recent burst from the SFHs predicted by \textsc{Bagpipes} for all of our galaxies in redshift intervals of 0.1. We then calculate the super-color in each redshift interval such that the photometric bands cover the same wavelengths as those at the galaxy's observed spectroscopic redshift. We assume the observed dust attenuation of our galaxies does not change pre- and post-quenching when constructing these evolutionary tracks. The visibility times depend on post-quenching dust, as galaxies are typically already quenched once they enter the PSB selection region. We expect dust clearing following a starburst to happen quickly, as the dust emission measured from far-IR bands in super-color selected PSBs is well below that of star-forming galaxies (Rowlands et al. in prep). Thus, we do not expect the visibility times to be affected much. However, if there is more dust in the galaxy at the time of quenching than there is at the time of observation, we expect the galaxies to have lower super-color 1 and super-color 2 values (i.e. moving to the lower left of the diagram). This may potentially decrease the derived visibility times we quote. If dust continues to clear in PSBs throughout the quenching process, then it may take longer for the galaxy to transition to the quiescent region, and thus increase the visibility time. We show the median ``super-color tracks" of two of the galaxies in our sample (EXCELS -- 94982 and EXCELS -- 55414) in Figure~\ref{Fig::sc_track}. The post-starburst galaxy EXCELS -- 94982 currently resides in the star-forming region of the super-color diagram, and after the burst of star formation begins climbing up towards the top of the post-starburst region. As star formation is nearly completely quenched, the galaxy moves towards the quiescent region of the diagram. The red star indicates the current position of the galaxy in super-color space, and the points below this redshift show the expected super-color values assuming no rejuvenation event occurs. In contrast, three of the quiescent galaxies in our sample (such as EXCELS -- 55414) never enter the post-starburst region of the diagram.

\begin{figure}
    \centering	
    \includegraphics[width=\linewidth]{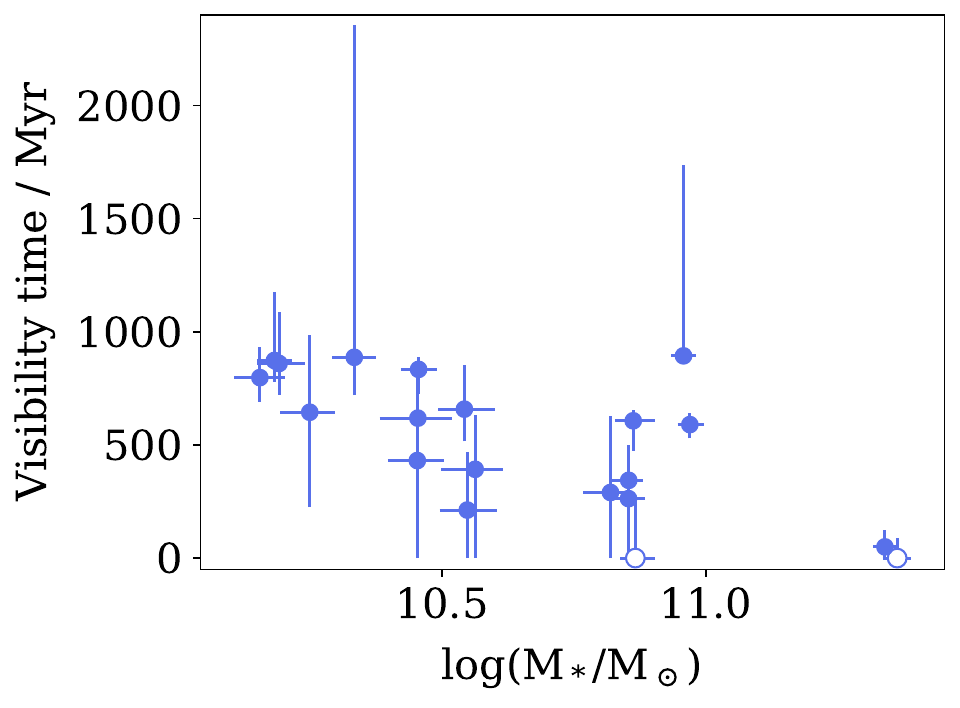}
    \caption{The visibility time of galaxies spent in the post-starburst region of the super-color diagram versus stellar mass. Open circles indicate galaxies that have a median visibility time of zero, but enter the PSB region within the uncertainty. We find a median visibility time of 612 Myr.}
    \label{Fig::vist_mass}
\end{figure}

Figure~\ref{Fig::vist_mass} shows the visibility time of post-starburst galaxies, defined as the amount of time spent in the post-starburst selection region in the super-color diagram, versus stellar mass. The median visibility time in our sample is 612 Myr, excluding six sources that never enter the PSB region and have a median visibility timescale of zero. We compare this timescale to those found in other studies in Section~\ref{sec:dis_vistime}. We perform a linear least-squares regression between the visibility time and stellar mass and find that the Pearson correlation coefficient is $-0.42^{+0.08}_{-0.09}$ with a p-value of ${0.08^{+0.08}_{-0.05}}$ (corresponding to a $1.8\sigma$ significance), suggesting that there is a tentative trend. 

\subsection{Line diagnostics}

\begin{figure}[h!]
    \centering	
    \includegraphics[width=\columnwidth]{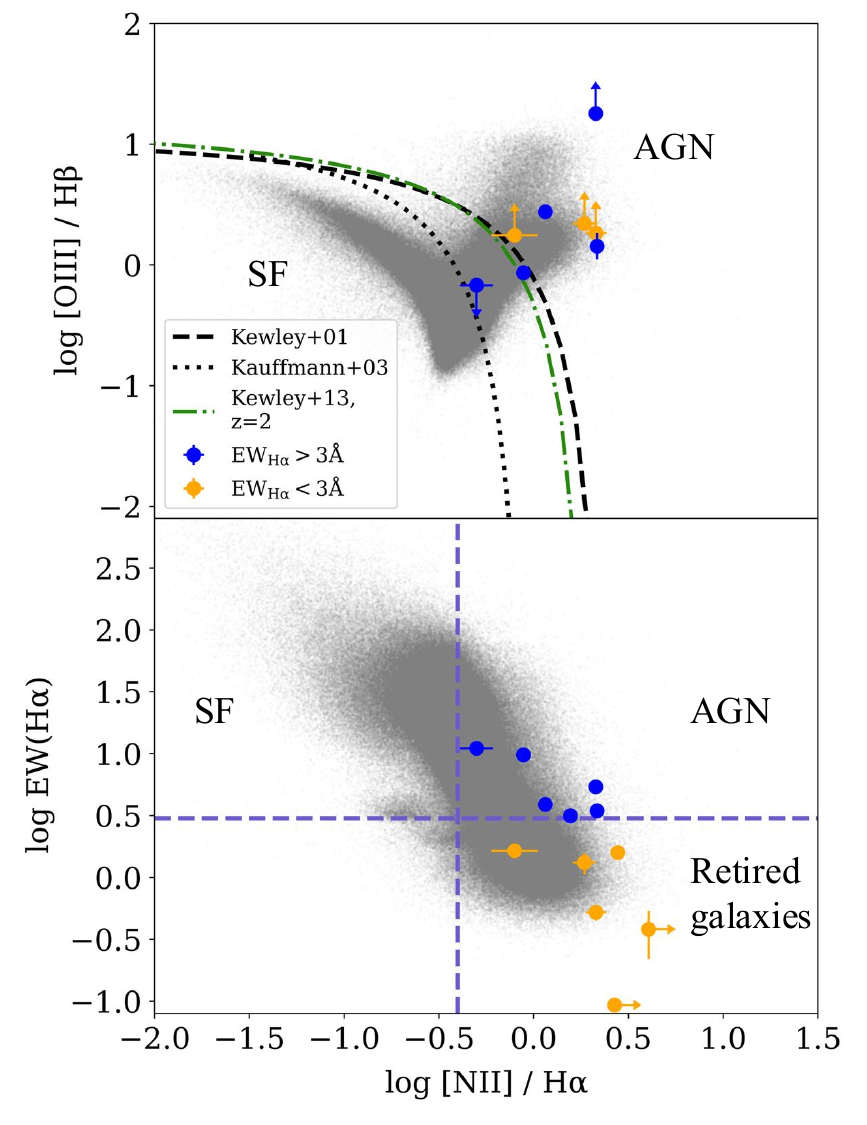}
    \caption{\textit{Top:} BPT diagram for our sample of post-starburst and quiescent galaxies. Galaxies with an EW$_{H\alpha}>3\textrm{\AA}$ are shown in blue and galaxies with an EW$_{H\alpha}<3\textrm{\AA}$ are shown in orange. The gray points show the distribution of local galaxies from SDSS \citep{Kauffmann2003b, Brinchmann2004, Tremonti2004}. \textit{Bottom:} WHaN diagram for our sample of post-starburst and quiescent galaxies. The lavender dashed lines indicate the separation between AGN, star-forming galaxies, and retired galaxies proposed by \citet{CF2011}.}
    \label{Fig::bpt}
\end{figure}

Here we explore the ionized gas in our sample of quenching and quenched galaxies. 
In the top panel of Figure~\ref{Fig::bpt}, we show the BPT diagram \citep{BPT1981} of the [NII]/H$\alpha$ versus [OIII]/H$\beta$ line ratios. There are eight galaxies with wavelength coverage of all four emission lines, where at least one emission line in each ratio is detected at $>3\sigma$. We include upper and lower detection limits when only one line is detected in either of the line ratios. We show the theoretical limit separating AGNs and starbursts from \cite{Kewley2001}, as well as the redshift-dependent classification line proposed by \cite{Kewley2013}. We also show the star formation limit derived by \cite{Kauffmann2003} based on a sample of nearby galaxies from the Sloan Digital Sky Survey. The majority of galaxies with robust line ratios fall into the AGN region of the BPT diagram. 
We also plot the distribution of local galaxies based on the DR8 MPA-JHU measurements\footnote{https://www.sdss4.org/dr17/spectro/galaxy\_mpajhu/} from the SDSS survey \citep{Kauffmann2003b, Brinchmann2004, Tremonti2004}. Our sample is clearly offset from the local star-forming sequence, with higher values of [NII]/H$\alpha$.

We show the WHaN diagnostic diagram proposed by \cite{CF2010, CF2011} in the bottom panel of Figure~\ref{Fig::bpt}. This diagram is useful in distinguishing ionization from weak AGN and `retired galaxies' (which are not forming stars and are ionized by hot evolved stars) in the absence of [OIII] and H$\beta$ \citep{CF2011}. The criterion $\log ($[NII]/H$\alpha) >-0.4$ is sometimes used to indicate non-stellar photoionization. Thirteen of the fifteen galaxies in our sample satisfy this condition (with two lacking robust detections for either line). However, as shown in the bottom panel of Figure~\ref{Fig::bpt}, about half of the galaxies with detections have an EW$_{H\alpha}<3\textrm{\AA}$, which \cite{CF2011} suggest are likely quenched galaxies with older stellar populations. We emphasize that this limit is likely conservative, given that this limit was designed for the local Universe and stellar populations tend to be younger at higher redshifts. The galaxies above (below) this limit are shown in blue (orange) in Figure~\ref{Fig::bpt}. 
Combined with classifications from the BPT diagram described above, four of the eight galaxies with robust line detections meet the AGN criteria in both diagnostics.
These four galaxies are on the higher mass end of the sample ($\log \textrm{M}_*/\textrm{M}_\odot > 10.8$) and have higher star formation rates. The measured BPT and WHaN diagnostics are summarized in Table~\ref{tab_line_diag}. We will further investigate whether there is AGN activity in Section~\ref{sec:agn}.

\begin{table*}
\centering
\begin{tabular}{llll}
\hline
 EXCELS ID   & EW H$\alpha$ ($\text{\AA}$)   & log([NII]/H$\alpha$)   & log([OIII]/H$\beta$)   \\
\hline
 $101499$    & $11.03^{+0.3}_{-0.3}$         & $-0.301 \pm 0.086$     & $<-0.168$              \\
 $93227$     & $0.52^{+0.07}_{-0.07}$        & $0.329 \pm 0.055$      & $>0.264$               \\
 $91726$     & $9.78^{+0.1}_{-0.1}$          & $-0.053 \pm 0.015$     & $-0.066 \pm 0.060$     \\
 $114633$    & --                            & --                     & $>0.176$               \\
 $123852$    & --                            & --                     & $>0.573$               \\
 $97765$     & --                            & --                     & not detected           \\
 $127460$    & $3.45^{+0.1}_{-0.1}$          & $0.334 \pm 0.010$      & $0.155 \pm 0.109$      \\
 $98447$     & $0.00^{+0}_{-0}$              & $>1.734$               & not detected           \\
 $114929$    & --                            & --                     & not detected           \\
 $114127$    & $1.32^{+0.3}_{-0.3}$          & $0.268 \pm 0.062$      & $>0.342$               \\
 $69069$     & --                            & --                     & $>0.306$               \\
 $54835$     & --                            & --                     & $>0.305$               \\
 $63832$     & --                            & --                     & $>0.376$               \\
 $43944$     & $3.14^{+0.3}_{-0.3}$          & $0.195 \pm 0.032$      & --                     \\
 $72186$     & --                            & --                     & not detected           \\
 $63089$     & $0.00^{+0}_{-0}$          & --                     & not detected           \\
 $65234$     & $0.38^{+0.2}_{-0.2}$          & $>0.607$               & not detected           \\
 $94982$     & $1.59^{+0.1}_{-0.1}$          & $0.445 \pm 0.015$      & not detected           \\
 $90591$     & $0.09^{+0.1}_{-0.09}$         & $>0.427$               & not detected           \\
 $122839$    & $3.88^{+0.1}_{-0.1}$          & $0.062 \pm 0.028$      & $0.440 \pm 0.046$      \\
 $90549$     & $0.40^{+0.2}_{-0.2}$          & not detected           & $>0.095$               \\
 $69759$     & $5.39^{+0.2}_{-0.2}$          & $0.328 \pm 0.013$      & $>1.255$               \\
 $60232$     & $0.00^{+0}_{-0}$              & not detected           & not detected           \\
 $55414$     & $1.64^{+0.1}_{-0.1}$          & $-0.100 \pm 0.122$     & $>0.246$               \\
\hline
\end{tabular}
\caption{Measurements for fitted emission lines. The equivalent width of H$\alpha$ is defined as positive for emission. The line ratios log([NII]/H$\alpha$) and log([OIII]/H$\beta$) are labeled as `not detected' if both emission lines are undetected at $>3\sigma$. Dashes indicate the the lines were not observed.}
\label{tab_line_diag}
\end{table*}

\section{Discussion}
\label{sec:dis}

In this paper, we present a sample of $z\sim 2$ quenching and quenched galaxies. In this section, we explore the visibility timescales of PSBs, their possible progenitors, and discuss the implications of the quenching timescales of these galaxies. Finally, we discuss the role AGN may be playing in quenching at cosmic noon. We summarize the relevant parameters discussed in this section in Table~\ref{tab_props}.

\begin{table*}
\centering
\begin{tabular}{llllll}
\hline
 EXCELS ID   & SFR$_{\text{max}} (\textrm{M}_\odot / \text{yr})$   & $z_{\text{SFR}_{\text{max}}}$   & Visibility time (Myr)   & $\tau_\text{q1}$ (Gyr)   & $t_{50} - t_{90}$ (Gyr)   \\
\hline
 $101499$    & $16^{+14}_{-4}$                                     & $4.0^{+5}_{-2}$                 & $644^{+340}_{-419}$     & $2.59^{+1}_{-1}$         & $1.71^{+0.6}_{-0.5}$      \\
 $114127$    & $295^{+358}_{-181}$                                 & $4.9^{+2}_{-0.9}$               & $392^{+242}_{-392}$     & $0.60^{+0.2}_{-0.5}$     & $0.31^{+0.2}_{-0.2}$      \\
 $114633$    & $1190^{+1361}_{-710}$                               & $3.5^{+0.2}_{-0.2}$             & $658^{+196}_{-142}$     & $0.06^{+0.1}_{-0.04}$    & $0.04^{+0.1}_{-0.03}$     \\
 $114929$    & $674^{+5708}_{-375}$                                & $5.4^{+13.0}_{-1.2}$            & $0^{+0}_{-0}$           & $0.49^{+0.7}_{-0.4}$     & $0.39^{+0.4}_{-0.3}$      \\
 $122839$    & $241^{+66}_{-112}$                                  & $2.4^{+0.2}_{-0.2}$             & $0^{+260}_{-0}$         & $1.58^{+0.2}_{-0.4}$     & $0.66^{+0.1}_{-0.1}$      \\
 $123852$    & $278^{+224}_{-141}$                                 & $8.4^{+10.4}_{-2.5}$            & $290^{+338}_{-290}$     & $0.85^{+0.2}_{-0.4}$     & $0.42^{+0.1}_{-0.2}$      \\
 $127460$    & $383^{+672}_{-170}$                                 & $3.6^{+0.4}_{-0.2}$             & $263^{+90}_{-263}$      & $0.34^{+0.3}_{-0.2}$     & $0.14^{+0.1}_{-0.1}$      \\
 $43944$     & $265^{+46}_{-114}$                                  & $4.0^{+1}_{-0.4}$               & $0^{+0}_{-0}$           & $0.72^{+0.1}_{-0.1}$     & $0.31^{+0.08}_{-0.05}$    \\
 $54835$     & $576^{+742}_{-255}$                                 & $3.1^{+0.2}_{-0.2}$             & $0^{+0}_{-0}$           & $0.35^{+0.2}_{-0.3}$     & $0.16^{+0.1}_{-0.09}$     \\
 $55414$     & $508^{+217}_{-258}$                                 & $2.4^{+0.3}_{-0.2}$             & $0^{+0}_{-0}$           & $0.87^{+0.2}_{-0.2}$     & $0.39^{+0.3}_{-0.06}$     \\
 $60232$     & $130^{+281}_{-98}$                                  & $1.8^{+0.07}_{-0.1}$            & $860^{+229}_{-138}$     & $0.11^{+0.3}_{-0.07}$    & $0.60^{+0.8}_{-0.6}$      \\
 $63089$     & $358^{+693}_{-182}$                                 & $2.7^{+0.2}_{-0.2}$             & $212^{+257}_{-212}$     & $0.21^{+0.3}_{-0.2}$     & $0.21^{+0.4}_{-0.2}$      \\
 $63832$     & $2042^{+2649}_{-1112}$                              & $3.2^{+0.2}_{-0.1}$             & $607^{+47}_{-131}$      & $0.08^{+0.2}_{-0.05}$    & $0.05^{+0.1}_{-0.03}$     \\
 $65234$     & $241^{+253}_{-111}$                                 & $2.6^{+0.08}_{-0.06}$           & $873^{+302}_{-95}$      & $0.08^{+0.1}_{-0.05}$    & $0.19^{+0.2}_{-0.2}$      \\
 $69069$     & $331^{+470}_{-155}$                                 & $2.6^{+0.1}_{-0.09}$            & $887^{+1468}_{-168}$    & $0.12^{+0.3}_{-0.08}$    & $0.12^{+0.2}_{-0.1}$      \\
 $69759$     & $317^{+204}_{-81}$                                  & $6.9^{+8}_{-3}$                 & $0^{+88}_{-0}$          & $1.71^{+0.6}_{-0.7}$     & $0.78^{+0.3}_{-0.3}$      \\
 $72186$     & $4695^{+3947}_{-2758}$                              & $3.5^{+0.1}_{-0.1}$             & $894^{+845}_{-29}$      & $0.06^{+0.1}_{-0.04}$    & $0.02^{+0.06}_{-0.01}$    \\
 $90549$     & $274^{+520}_{-197}$                                 & $1.8^{+0.08}_{-0.1}$            & $618^{+194}_{-195}$     & $0.12^{+0.3}_{-0.07}$    & $0.48^{+0.8}_{-0.4}$      \\
 $90591$     & $138^{+156}_{-69}$                                  & $1.4^{+0.06}_{-0.06}$           & $798^{+135}_{-106}$     & $0.15^{+0.3}_{-0.1}$     & $0.23^{+0.9}_{-0.2}$      \\
 $91726$     & $198^{+286}_{-95}$                                  & $2.1^{+0.4}_{-0.2}$             & $344^{+156}_{-344}$     & $0.78^{+0.7}_{-0.5}$     & $0.31^{+0.3}_{-0.2}$      \\
 $93227$     & $179^{+344}_{-112}$                                 & $1.5^{+0.09}_{-0.08}$           & $431^{+311}_{-431}$     & $0.24^{+0.5}_{-0.2}$     & $0.62^{+0.9}_{-0.6}$      \\
 $94982$     & $1004^{+94}_{-164}$                                 & $2.4^{+0.05}_{-0.05}$           & $590^{+51}_{-61}$       & $0.51^{+0.05}_{-0.06}$   & $0.17^{+0.02}_{-0.02}$    \\
 $97765$     & $623^{+1096}_{-333}$                                & $3.1^{+0.1}_{-0.1}$             & $834^{+54}_{-108}$      & $0.11^{+0.2}_{-0.08}$    & $0.06^{+0.1}_{-0.05}$     \\
 $98447$     & $1867^{+1405}_{-631}$                               & $3.4^{+0.08}_{-0.05}$           & $50^{+73}_{-50}$        & $0.21^{+0.1}_{-0.1}$     & $0.09^{+0.04}_{-0.04}$    \\
\hline
\end{tabular}
\caption{Measured properties of our galaxy sample, with values given as posterior medians and errors derived from the 16th and 84th percentiles of the BAGPIPES posterior distributions. The columns are (1) EXCELS ID number, (2) the maximum historical star formation rate, (3) the redshift this star formation rate is achieved, (4) the time that galaxies remain in the super-color post-starburst selection region, (5) the time from peak star formation to sSFR$= 0.2/t_{\text{H}}$, where $t_{\text{H}}$ is the Universe's age at the time of quenching, (6) the time to grow from 50\% to 90\% of the total stellar mass.
}
\label{tab_props}
\end{table*}

\subsection{Visibility times}
\label{sec:dis_vistime}
The visibility timescales of PSBs can be used to constrain the importance of PSBs for the growth of the quiescent galaxy population \citep[e.g.][]{Wild16, Taylor23}. The visibility times found in this sample (see Section~\ref{sec:res_SCs}) have a similar range to the post-starburst visibility times found by \cite{Wild2020} with a sample of galaxies at $0.5<z<1.3$, and show no evidence for evolution in visibility times up to $z\sim 3$. This might suggest that the same physical mechanism(s) are responsible for quenching star formation at $z\sim 1$ and $z\sim 2$. Additionally, we find a tentative trend between the visibility time and stellar mass (a $1.8\sigma$ trend, see Section~\ref{sec:res_SCs}), although this is over a narrow range of stellar masses and we are limited by a small sample. In other papers that also define PSBs using super-colors, the PSB visibility timescale is estimated by different methods.
de Lisle et al. (in prep) calculate visibility times by comparing mass functions of post-starburst and quiescent galaxies at different epochs and find that visibility times get shorter with increasing stellar mass between $10^9\text{M}_\odot$ and $10^{10} \text{M}_\odot$, but with no clear trend at higher masses. Harrold et al. (in prep) calculate the visibility times of PSBs using semi-analytic models, and find a tentative trend with visibility times getting shorter with increasing stellar mass between $10^{9.5}\text{M}_\odot$ and $10^{11.5} \text{M}_\odot$. The tentative trend in the visibility times with stellar mass may be due to high- and low-mass galaxies undergoing different quenching processes that act on different timescales \citep[e.g.][]{Maltby18}. Low-mass galaxies, for example, may be more likely to be quenched by environmental processes, which may result in longer visibility timescales. AGN feedback is found to be a primary pathway for quenching massive galaxies at high-redshift in simulations \citep[e.g.][]{Chittenden2025}, which may lead to shorter visibility times.

\cite{Taylor23} found that PSBs may account for about 50\% of the growth of the massive ($\log \textrm{M}_*/\textrm{M}_\odot > 10.7$) quiescent galaxy population, and potentially all the growth at lower masses, at $z\sim2$ assuming a visibility time of 500 Myr. If the visibility times are longer, the contribution of PSBs to the growth of the quiescent galaxy population would be lower, making them a less dominant pathway. Taking the 20\% longer visibility time that we measure from our sample decreases this estimate to PSBs contributing about 40\% to the growth of the quiescent galaxy population. However, when splitting by mass, the median visibility time of $\sim340$ Myr at $\log \textrm{M}_*/\textrm{M}_\odot > 10.7$ increases the estimated contribution of PSBs to the growth of the massive quiescent galaxy population to $\sim73\%$. The longer median visibility time of $\sim660$ Myr at $\log \textrm{M}_*/\textrm{M}_\odot < 10.7$ allows for some of the growth of the quiescent sequence via galaxies that do not undergo a post-starburst phase. Thus, PSBs are an important pathway in the growth of quiescent galaxies at $z\sim 2$.

\subsection{Progenitors of quenched galaxies at cosmic noon}
\begin{figure}
    \centering	
    \includegraphics[width=\linewidth]{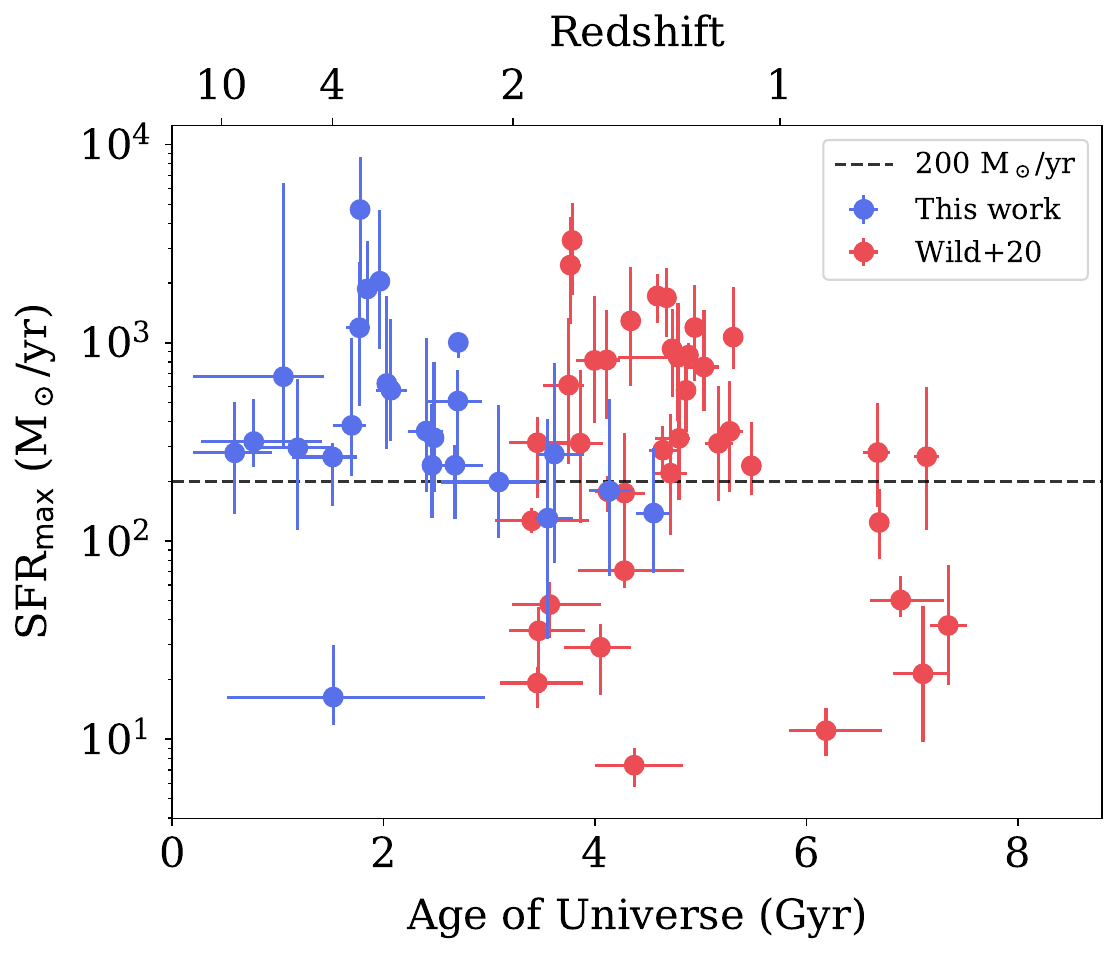}
    \caption{The peak historical star formation rate as a function of the age of the Universe when this occurred for galaxies in this sample and from the $z\sim1$ \cite{Wild2020} sample. The dashed black line indicates a SFR of 200 $\text{M}_\odot \text{yr}^{-1}$, which we consider as a lower limit for SMGs.}
    \label{Fig::sfrmax}
\end{figure}

Submillimeter galaxies (SMGs) have been suggested as possible progenitors of post-starburst galaxies \citep[e.g.][]{Almaini2017, Wild2020, Wilkinson21}. To investigate this possibility, we examine the peak historical SFR derived from \textsc{Bagpipes} for our sample as a function of the redshift at which this occurs in Figure~\ref{Fig::sfrmax}. The range of historical maximum SFRs (16-4695 $\text{M}_\odot \text{yr}^{-1}$) is consistent with those found in \cite{Wild2020} for $z\sim 1$ PSBs. Note that for the historical maximum SFRs of the \cite{Wild2020} sample we consider the peak of the recent starburst, which is slightly lower for a few galaxies that did not have a strong recent starburst.
The range of SFRs of SMGs found by \cite{Swinbank2014} is $20-1030$ $\text{M}_\odot \text{yr}^{-1}$, which is consistent with the SFRs achieved by all but one of the galaxies in our sample. SMGs have been found to have stellar masses and space densities consistent with them being the progenitors of post-starburst galaxies \citep{Wilkinson21}. 

To investigate whether the galaxies in our sample are the descendants of SMGs, we consider the sample of SMGs presented in \citet{Gillman2024}. There are seventeen galaxies in our sample that have SFRs that exceed $200$ $\text{M}_\odot \text{yr}^{-1}$ in the redshift range of the \citet{Gillman2024} sample ($1.4<z<4.6$). We assume that our lower limit of $200$ $\text{M}_\odot \text{yr}^{-1}$ roughly corresponds to the ALMA 870 \textmu m $S_{870 \textmu m} = 2$ mJy \citep[e.g.][]{Hayward2011}. Considering the number density of sources in the \citet{Gillman2024} sample within the UDS field, and the number density of sources in our sample likely to have undergone an SMG phase (assuming our sample is representative of the parent sample in the EXCELS footprint), we find that our sample has a density $\sim 7\times$ higher than that of the SMGs. However, SMGs are visible for much shorter time scales. Based on our predicted SFHs, our galaxies only exceed SFRs of $200$ $\text{M}_\odot \text{yr}^{-1}$ for around $69^{+27}_{-21}$ Myr, which is about $\sim 7$ times shorter than the time our galaxies have been visible as post-starburst or quiescent. Thus, based on these number densities and visibility times, SMGs are likely progenitors of the galaxies in our sample.

Additionally, post-starburst galaxies have been found to be compact at $z>1$ \citep[e.g.][]{Almaini2017, Ji2024}, consistent with the compact sub-millimeter sizes from ALMA, although the rest-frame optical sizes of SMGs are $\sim2-4$ times larger \citep{Simpson2015, Lang2019}. All of the galaxies in our sample exhibit extremely compact morphologies with a median effective radius of $\sim 1.1$ kpc (Maltby et al. in prep), further supporting the idea that these galaxies may be the descendants of SMGs.

\subsection{Quenching timescales}
\begin{figure}
    \centering	
\includegraphics[width=\linewidth]{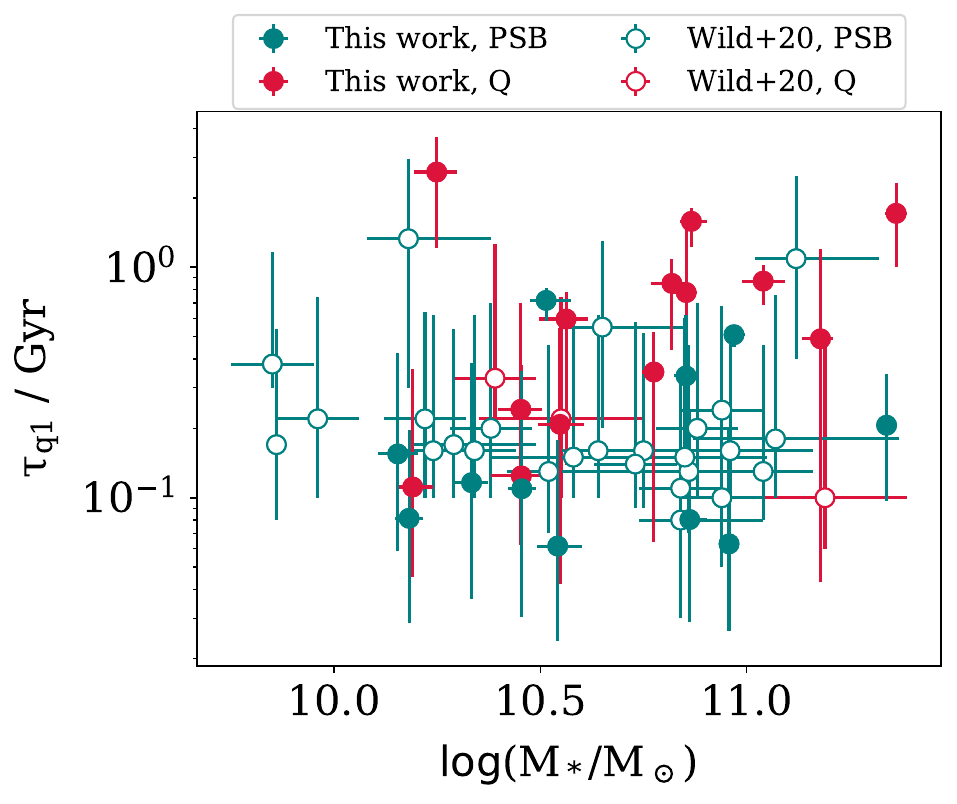}
    \caption{Quenching timescale as a function of stellar mass found for this sample of galaxies and those from \cite{Wild2020}. The points are colored according to their super-color classification. We include the few galaxies classified as star-forming (but near the quiescent boundary) in the quiescent sample.}
    \label{Fig::tau_q1}
\end{figure}

Different quenching timescales can be used to probe various quenching mechanisms. Figure~\ref{Fig::tau_q1} shows the quenching timescale $\tau_{\text{q1}}$ as a function of stellar mass. The quenching timescale $\tau_{\text{q1}}$ is defined as the amount of time from peak star formation to sSFR$= 0.2/t_{\text{H}}$, where $t_{\text{H}}$ denotes the Universe's age at the time of quenching. There are fifteen galaxies in our sample that quench within 500 Myr, six galaxies that quench between 500 Myr and 1 Gyr, and three galaxies that take longer than 1 Gyr to fully quench. The galaxies with the highest maximum SFRs tend to have the fastest quenching times, possibly due to winds driven by star formation, star formation fueling an AGN, or gas depletion. 

A large fraction of our sample has quenching times longer than those in \cite{Wild2020}, which is due to our sample including quiescent galaxies. If PSBs become quiescent, then the quiescent sample should contain some galaxies with rapid quenching times, although the errors are expected to be larger as the star formation burst occurred longer ago. If there are some quiescent galaxies that do not have short quenching times, that allows for the other approximately 50\% of quenched galaxies that do not go through a PSB phase, as predicted by \citet{Taylor23}. Of the nine quiescent galaxies in our sample, six have undergone a PSB phase in the past (based on the ``super-color tracks" described in Section~\ref{sec:res_SCs}) of which four have quenched within 500 Myr. The other three quiescent galaxies never had a PSB phase.

We also consider a quenching timescale that is not dependent on the sSFR. The quenching time $t_{50} - t_{90}$ is defined as the time it takes a galaxy to grow from 50\% to 90\% of its total formed stellar mass. We see a range of quenching timescales, with faster quenching timescales more common at higher redshift, although this is partly due to the limited time available for galaxies to form and gradually quench when the Universe was so young. We note that the $t_{50} - t_{90}$ statistic is only appropriate as a quenching time measurement for galaxies which undergo a single episode of star formation, which is not consistent for galaxies with an extended SFH prior to the most recent starburst. \cite{Kimmig2025} find that most quenched galaxies in the \textsc{Magneticum} simulations have quenching timescales of $t_{50} - t_{90}<170$ Myr. There are eight galaxies in our sample with $t_{50} - t_{90}<170$ Myr. However, many of the timescales found in this work and in \cite{Wild2020} are significantly longer, and are sometimes even longer than 1 Gyr. While \textsc{Magneticum} is able to reproduce the observed number densities of quiescent galaxies at $z\sim3-4$, they over-predict the number density of $z\sim2$ quiescent galaxies likely due to an overly efficient AGN feedback model \citep{Lagos2025}. This may explain the very fast quenching times found by \cite{Kimmig2025}.

The timescales we find in this work are broadly consistent with the wide range of quenching timescales found in other galaxy populations. \cite{Tacchella2022} find a wide range of quenching timescales (0-5 Gyr) based on the time it takes to transition from $\text{sSFR} < 1/[20 t_{\text{universe}}]$ to $\text{sSFR} < 1/[3 t_{\text{universe}}]$ in a sample of massive quiescent galaxies at $z\sim 0.8$. Samples of young quiescent galaxies at $z\sim 1.5$ that have undergone a major starburst have quenching times around $\tau_{\text{q1}} \sim 300$ Myr \citep{Park2023}, which is longer than the median quenching time of $\sim116$ Myr for post-starburst galaxies in this sample. 

The diversity of quenching timescales we see in this sample suggest that these galaxies are being quenched by different processes. Studying the cold gas reservoirs in conjunction with the SFHs of quenching galaxies may shed light on quenching mechanisms. A simple explanation for slow quenching is gas starvation. In the local Universe, cutting off the supply of cold gas to galaxies may result in quenching timescales on the order of several Gyr \citep{Schawinski2014, Peng2015}. There is evidence that AGN activity may shorten quenching times \citep{Wright2019}, so a plausible explanation for our rapidly quenched galaxies ($\tau_{\text{q1}}<500$) is gas exhaustion from a major starburst coupled with AGN feedback. \cite{Belli2019} find that galaxies that quench rapidly likely undergo a compaction event (i.e., a rapid buildup of a dense stellar core) due to their compact sizes and spheroidal morphologies followed by quenching due to gas exhaustion and stellar feedback. Compaction events may be triggered by processes such as major or minor mergers, counter-rotating streams, or tidal-compressions that result in an inflow episode \citep{Zolotov2015}. Our sample also has compact spheroidal galaxies with a median effective radius of $1.1$ kpc and a median Sérsic index of $n=3.6$ (Maltby et al. in prep), suggesting many of our galaxies have undergone compaction events prior to quenching.

We do not see a trend with the quenching time scales and stellar mass, possibly due to our small sample size. Looking at galaxies that are passive at $z=0$ in the EAGLE simulation, \citet{Wright2019} find that there is a negative correlation between quenching timescales and mass at $\log \textrm{M}_*/\textrm{M}_\odot > 9.7$, although the environment of the galaxy likely also plays a role. \citet{Peng2010} proposes that there are two distinct processes which can quench galaxies up to $z\sim1$, “mass quenching” and “environment quenching”. While we defer a detailed analysis of the environments of our sample of galaxies to a future study, \citet{Taylor23} find that at a given stellar mass and redshift, denser environments have higher quenching rates out to at least $z\sim 2$. However, since mass quenching is thought to be more important at $\log \textrm{M}_*/\textrm{M}_\odot > 10.2$, possibly due to feedback mechanisms linked to star formation or AGN activity \citep{Peng2010}, the environment alone is unlikely to be the primary driver of quenching in our sample.

\subsection{The Presence of AGNs}
\label{sec:agn}

The role of AGN in quenching star formation in galaxies remains uncertain. Current simulations suggest that while gas consumption and stellar feedback cause the initial decline in the SFR, an additional mechanism such as AGN feedback must be invoked to completely quench the galaxy \citep{Zheng2020}. Additionally, recent cosmological simulations struggle to reproduce the properties of massive quiescent galaxies in the early universe and the AGN feedback model implemented changes the predicted number densities of quiescent galaxies and the quenching time-scales \citep{Lagos2025}. 
There are two main modes of AGN activity that can be important for quenching. The “quasar” mode is associated with high accretion rates and is often invoked to drive the initial quenching episode \citep{Hopkins2006}, while the “radio” mode is associated with low levels of accretion. The radio-mode feedback is thought to suppress cooling flows \citep{Croton2006}, and is therefore often referred to as a “maintenance” mode. However, there is some evidence from simulations that the consumption of dense ISM gas following a merger, coupled with morphological quenching, is sufficient to halt star formation without invoking AGN feedback \citep{Petersson2023}. Out of the five galaxies in our sample with reliable line ratios that satisfy $\text{EW}(\text{H}\alpha)> 3 \text{\AA}$, such that the ionization lines are likely not due to evolved stars \citep{CF2011}, four fall in the AGN region of the BPT diagram (see Figure~\ref{Fig::bpt}). The BPT diagram is not able to distinguish between shocks driven by starbursts and photoionization due to an AGN \citep{Kewley2019}. However, the low SFRs present in these systems suggests that if shocks are responsible for the high BPT emission line ratios, they are unlikely to be driven by stellar feedback. While merger-induced shocks may also reproduce the emission line ratios of AGN in the BPT diagram \citep{Medling2015}, we do not see evidence for mergers in the NIRCam imaging for any of our galaxies, although we note that merger signatures may disappear on short timescales, so features such as tidal tails might not be visible. It is therefore likely that the high emission line ratios for these galaxies are due to AGN activity.

Several studies have also found evidence for AGN activity using line ratio diagnostics in quenching or quenched galaxies at cosmic noon \citep[e.g.][]{Belli2024, DEugenio2024, Bugiani2024} and in the local Universe \citep[e.g.][]{Yesuf2014, Pawlik2018}. Fast outflows ($\sim 1000 \textrm{ km s}^{-1}$) at $z>1$ seen in quenched galaxies have been interpreted as being driven by AGN activity, as these galaxies have low current star formation rates and the outflows are detected long ($>0.6$ Gyr) after the peak in star formation \citep{Taylor2024}.
In a sample of quiescent galaxies at cosmic noon from the Blue Jay survey, \cite{Bugiani2024} found that 60\% of the galaxies reliably classified from line ratios show signs of hosting an AGN. This is consistent with the 50\% of galaxies in our sample with reliable classifications showing evidence of AGN activity. Similar detection fractions have also been found in samples of quiescent galaxies up to $z\sim5$ \citep{Baker2025, Stevenson2025}. None of the galaxies in our sample are detected in the X-ray in \textit{Chandra} by the X-UDS survey \citep[PI: G. Hasinger; ][]{Kocevski2018}, and only one galaxy (EXCELS -- 94982) is detected in the radio (Patil et al. in prep). \cite{Bugiani2024} similarly found that none of the quiescent galaxies in their sample are detected in the X-rays or radio. \citet{MartinezMarin2024} model the dust in the torus to obtain the fraction of IR luminosity due to an AGN for a sample of 22 massive galaxies at $3<z<4$ and find that many galaxies have a highly-dominant AGN. However, none of our 24 sources are detected at 24 \textmu m with the \emph{Spitzer} UDS Legacy Program (SpUDS, PI:Dunlop where the data is available in IRSA \citep{https://doi.org/10.26131/irsa403}), implying a lack of AGN-heated dust. Our sources with MIRI F770W coverage are detected and 3 of the 7 sources with MIRI F1800W coverage in our sample are detected, but none of the sources are significantly red in the F770W-F1800W color (Hewitt, private communication). While this cannot rule out any AGN activity, it does suggest that if there is an AGN it is likely not highly accreting. The one exception is a dusty galaxy, EXCELS – 114929, which has an enhanced F770W-F1800W color, although we note that this color selection is less useful at $z>2$ and therefore does not necessarily imply any AGN activity.

In the local Universe, \cite{Lanz2022} find that half of their sample of 12 PSBs show evidence of AGN activity with low X-ray luminosities, and suggest that AGN may be `along for the ride', having been fueled by the same processes that caused the starburst and/or quenching episode. At cosmic noon ($0.5<z<3$), there is no evidence for enhanced X-ray AGN activity in PSBs, with the observed weak AGN activity consistent with the low star formation rates \citep{Almaini2025}. Patil et al. (in prep) similarly find low detection rates of PSBs at cosmic noon in the radio, and a stacking analysis reveals very weak AGN activity or low-level star formation in the most massive post-starburst galaxies. This is consistent with the lack of radio and X-ray detections in quenching galaxies at cosmic noon, despite the high AGN fractions inferred from line ratios.
It is also possible that the galaxies in our sample are classified as AGN due to their higher mass, which makes it easier to detect weak AGN. Indeed, of the galaxies in our sample with both line ratios available, the more massive galaxies with higher star formation rates tend to be classified as AGN based on line ratio diagnostics. Careful consideration of the sample selection, particularly the mass limits, is essential, as biases make it difficult to draw definitive conclusions.

Placing constraints on the gas reservoirs of quenching galaxies is important for constraining the physical mechanism(s) involved in quenching. The discovery of molecular gas reservoirs in post-starburst galaxies challenges the notion that quenching occurs solely due to gas starvation \citep[e.g.][]{Rowlands2015, Bezanson2022}. The recent detection of molecular gas in a quiescent $z\sim3$ galaxy \citep[][]{Umehata2025} further suggests that quenching galaxies at cosmic noon may be quenching star formation due to additional processes beyond gas depletion. However, even if reservoirs of cold gas remain in our sample, occasional bursts of AGN activity may suppress star formation via ejective or maintenance mode feedback. Using cosmological simulations, \cite{Kimmig2025} find that while AGN feedback is important in quenching galaxies, a rapid starburst followed by gas removal from AGN is the dominant quenching mechanism down to $z\sim 2$. This scenario seems plausible for the galaxies in our sample with strong starbursts followed by fast quenching times. 

The lack of X-ray and radio detections in our sample may be due to short AGN duty cycles, as the evidence for AGN activity based on line ratio diagnostics may reflect past AGN episodes \citep{French2023}. It is also possible that our sources are obscured or that the AGN are very weak, though still able to suppress star formation by disturbing the gas reservoir \citep{Smercina2022, Luo2022}.
AGN may play a crucial role in maintaining quiescence \citep[e.g.][]{Kimmig2025}, but AGN feedback does not seem to be the primary quenching driver, as our observations suggest that the driver is a combination of gas exhaustion and stellar feedback from a recent starburst.

\section{Conclusions}
\label{sec:conc}
We present a sample of 24 quenching and quenched massive ($\log M_*/M_\odot > 10$) galaxies at $1 < z < 3$ with medium-resolution NIRSpec spectroscopy from the EXCELS survey. We fit both the observed photometry and spectroscopy using \textsc{Bagpipes} to infer their star formation histories and emission line properties.
Our conclusions are summarized as follows:
\begin{enumerate}
    \item We confirm that the PCA photometric (super-color) selection of quiescent and post-starburst galaxies, first established by \citet{Wild14}, is robust up to redshifts of $z\sim3$ using EXCELS spectroscopy. Within the post-starburst region of the super-color diagram, we find a tentative trend with the super-color values and the burst age, with younger post-starburst galaxies having higher values of super-color 1 and super-color 2.
    \item We find that the median post-starburst visibility time is 612 Myr, ranging from 50 Myr to 894 Myr as shown in Figure~\ref{Fig::vist_mass}. Six out of nine of the quiescent galaxies in our sample are predicted to have entered the PSB region during their evolution, with the others quenching more slowly. PSBs are an important pathway for the growth of the quiescent galaxy population. Based on our predicted visibility times, PSBs potentially account for $\sim73\%$ of the growth of quiescent galaxies at the highest stellar masses ($\log \textrm{M}_*/\textrm{M}_\odot > 10.7$).
    \item We see a variety of SFHs, with the majority of the sample undergoing rapid quenching following a burst of star formation. The inferred maximum star formation rates of these galaxies (see Figure~\ref{Fig::sfrmax}) are consistent with many of them being the descendants of SMGs. 
    We find that the number densities of galaxies in our sample that went through an SMG phase between $1.4<z<4.6$ are consistent with the number density of SMGs, and predict that SMGs are visible for $\sim 70$ Myr based on our SFHs.
    \item In Figure~\ref{Fig::tau_q1}, we show that the quenching timescales in our sample span 0.06-1.75 Gyr. This wide range of quenching timescales suggests different quenching pathways may be at work.
    \item We find that 4 out of 8 galaxies with robust line detections in our sample show evidence for AGN activity based on rest-frame optical line diagnostics (see Figure~\ref{Fig::bpt}), despite none of these sources having X-ray or radio detections. This is consistent with other works at the same redshift \citep[e.g.][]{Bugiani2024}.
\end{enumerate}

There is likely a range of quenching mechanisms at $z\sim2$, resulting in the build-up of the quiescent galaxy population. We suggest that many of these galaxies may quench rapidly following a starburst due to stellar feedback and gas exhaustion, with AGN possibly playing a role in maintaining quiescence. Future work using larger samples of \textit{JWST}/NIRSpec observations will enable more robust tests of how post-starburst galaxy properties vary with burst age, providing deeper insights into the mechanisms driving quenching.

\section*{Acknowledgments}
We thank the anonymous referee for their thoughtful comments which has improved this work.
We would like to thank Celia Mulcahey and Decker French for helpful discussions. Based on observations with the NASA/ESA/CSA James Webb Space Telescope, obtained via the Mikulski Archive for Space Telescopes at the Space Telescope Science Institute, which is operated by the Association of Universities for Research in Astronomy, Incorporated. The specific observations analyzed in this work can be accessed at \dataset[doi: 10.17909/4r31-j678]{https://doi.org/10.17909/4r31-j678}. Support for Program number JWST-GO-03543.014 was provided through a grant from the STScI under NASA contract NAS5-03127. The VANDELS spectra used are available at http://vandels.inaf.it/. We gratefully acknowledge support from the NASA Astrophysics Data Analysis Program (ADAP) under grant 80NSSC23K0495 and the Maryland Space Grant. 
This work was carried out in part at the Advanced Research Computing at Hopkins (ARCH) core facility  (rockfish.jhu.edu), which is supported by the National Science Foundation (NSF) grant number OAC1920103. V.W. acknowledges support from STFC grant (ST/Y00275X/1) and Leverhulme Research Fellowship (RF-2024-589/4). 
J.O. acknowledges support from the Space Telescope Science Institute Director's Discretionary Research Fund grants D0101.90296 and D0101.90311. O.A. acknowledges the support from STFC grant ST/X006581/1. ACC, H-HL, SS and ET acknowledge support from a UKRI Frontier Research Guarantee Grant (PI Carnall; grant reference EP/Y037065/1). F. C acknowledges support from a UKRI Frontier Research Guarantee Grant (PI Cullen; grant reference EP/X021025/1). 


%

\facilities{JWST(NIRSpec), JWST(NIRCam), HST(ACS), JWST(MIRI), VLT}


\software{Astropy \citep{2013A&A...558A..33A,2018AJ....156..123A, Astropy2022},  Bagpipes \citep{Carnall2018, Carnall2019}, pPXF \citep{Cappellari2023}, Scipy \citep{2020SciPy}, Numpy \citep{numpy}, Matplotlib \citep{Matplotlib}, spectres \citep{Carnall2017}}



\appendix
\section{Spectra for sample of galaxies}
\label{appendix_a}
\begin{figure*}
    \centering	
\includegraphics[width=0.9\linewidth]{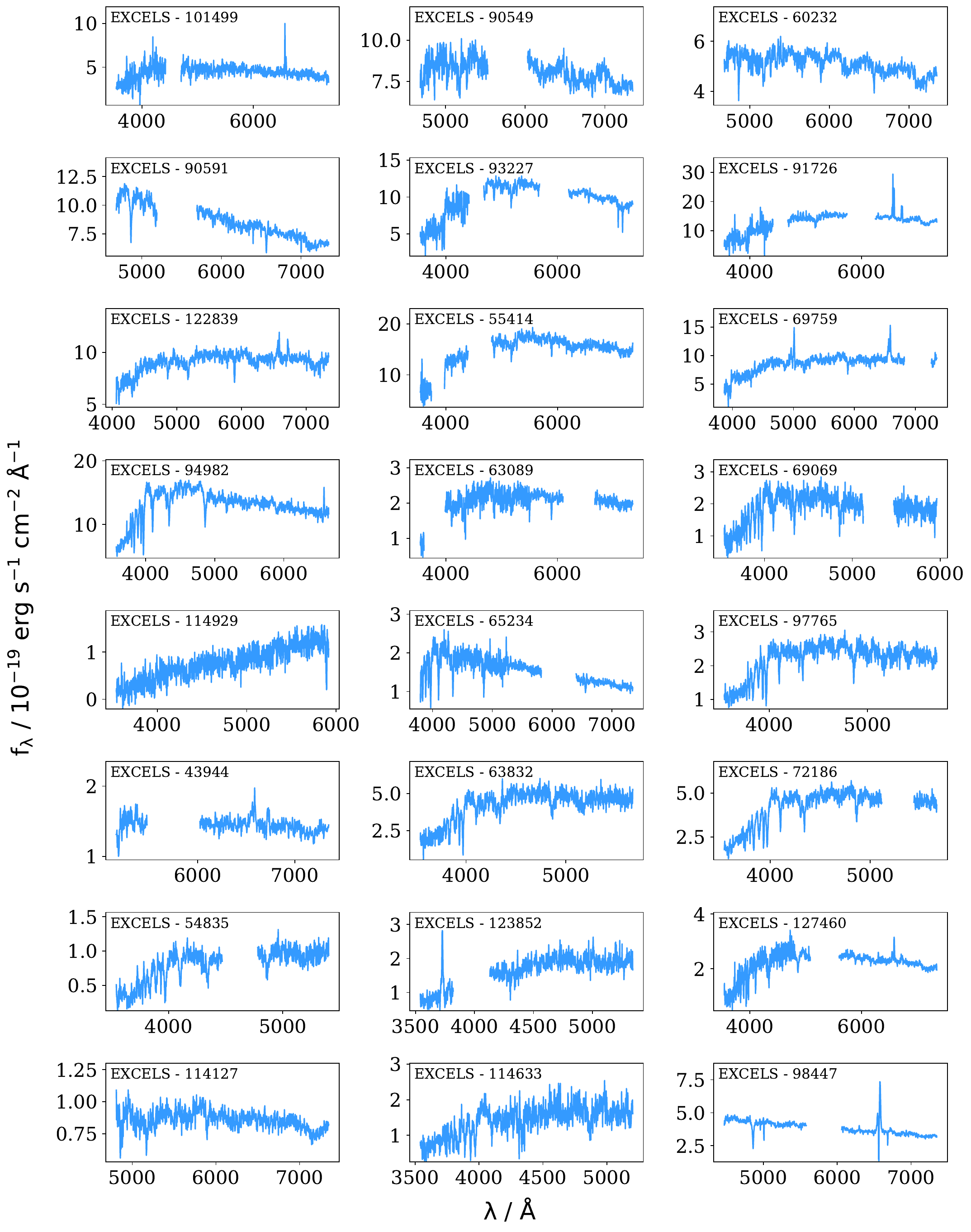}
    \caption{Rest-frame spectra for our sample of galaxies in the 3540 to $7350 \textrm{Å}$ range, sorted by redshift.}
    \label{Fig::all_spec}
\end{figure*}

In Figure~\ref{Fig::all_spec}, we show the rest-frame spectra for our sample of galaxies.


\bibliography{sample631}{}
\bibliographystyle{aasjournal}



\end{document}